\documentclass[12pt,english,floatfix,superscriptaddress,aps,prd,preprint,showkeys]{revtex4}
\usepackage{amsmath}
\usepackage{amssymb}
\usepackage{amsbsy}
\usepackage{amsfonts}
\usepackage{amsopn}
\usepackage{amstext}
\usepackage{graphicx}
\usepackage{amssymb}
\usepackage{amsfonts}
\usepackage{amsmath}
\usepackage{graphicx}
\usepackage[english]{babel}
\usepackage{color}
\usepackage{slashed}
\usepackage{esint}
\usepackage[dvips]{epsfig}
\usepackage[dvips]{graphicx}
\usepackage{float}
\usepackage{units}
\usepackage{textcomp}

\usepackage{slashed}
\usepackage{graphicx}
\usepackage{amssymb}
\usepackage{amsmath}
\usepackage{color}

\begin{document}



\title{Quantum information for a solitonic particle with hyperbolic interaction}


\author{A.R.P. Moreira}
\email{allan.moreira@fisica.ufc.br}
\affiliation{Universidade Federal do Cear\'a (UFC), Departamento de F\'isica,\\ Campus do Pici, Fortaleza - CE, C.P. 6030, 60455-760 - Brazil.}





\begin{abstract}

In this work, we analyze a particle with position-dependent mass, with solitonic mass distribution in a stationary quantum system, for the particular case of the BenDaniel-Duke ordering, in a hyperbolic barrier potential. The kinetic energy ordering of BenDaniel-Duke guarantees the hermiticity of the Hamiltonian operator. We find the analytical solutions of the Schrödinger equation and their respective quantized energies. In addition, we calculate the Shannon entropy and Fisher information for the solutions in the case of the lowest energy states of the system.

\end{abstract}


\maketitle
\section{Introduction}

Quantum Mechanics is considered one of the fundamental theories to describe the dynamic evolution and understand the wave behavior of matter, mainly in the aspect related to its structure, for analyzing in-depth the behavior of a particle in the microscopic medium. The quantum theory had a significant beginning around 1926, when Erwin Schrödinger formulated the equation that would become popularly known under his name \cite{Schr,Griffiths}. In 1927, Heisenberg deduced an uncertainty in the measurements regarding the position and moment of a particle in a microscopic medium, thus implying that the physically acceptable solutions of the Schrödinger equation were in their nature information system probabilistic analysis \cite{Griffiths}.

The mechanics proposed by Erwin Schrödinger, faced a new challenge with the theoretical advances in solid-state physics, where particles appear that behave unusually, with an effective mass that depends on the position \cite{Roos,Bastard,Weisbuch}. Since then, the number of researchers interested in working with quantum systems whose particle has a position-dependent mass (PDM) has increased. The interest of these researchers comes from the importance of this approach, which can describe some problems such as impurities in crystals \cite{Luttinger,Nabu}, study of electronic properties of semiconductor heterostructures \cite{Bastard,Pourali,Kasapoglu}, some related applications to the hermiticity of the Hamiltonian operator \cite{Mustafa}, to atomic and molecular physics \cite{Sever}, to supersymmetry \cite{Plastino}, to relativistic problems \cite{Almeida,Almeida1,Almeida2} and non-relativistic \cite{Cunha,Dong,Dong1}, to theoretical studies of Fisher's information calculation \cite{Falaye,Macedo2015,Lima2021} and Shannon entropy \cite{Lima2021,Sun2015,Dong2016,Navarro}.

Parallel to the advance of quantum theory to describe the behavior of a particle, whose mass depends on the position, information theories emerged. This theory began around 1948, when the mathematician and engineer Claude Shannon proposed one of the main concepts in the theory called entropy \cite{Shannon}, defined as a measure of how good is the propagation of information from a source to a receiver, that is, it is associated with the amount of “information” that is obtained with as little interference as possible \cite{Kripp,Zhou,Grasselli,Amigo}. Because this theory is linked to a probability density it could be applied to a quantum process, where the Shannon entropy $S_ {x}$ and $S_ {p}$, are respectively related to measuring the uncertainty of the particle location in the position space and in the momentum space \cite{Lima,Lima1}.

Another theory of great importance for communication and well before the concept of Shannon's theory is Fisher's information \cite{Fisher}. The interesting thing about this foundation is that it also works with probability density, which can be applied to a quantum system \cite{Shi,Arenas,Ikot}. In this same context, Fisher's information is also intrinsically related to the uncertainty of a measurement, in other words, Fisher's information is a way of measuring the amount of information that certain observable carries about a certain parameter with a related intrinsic probability \cite{Shi,Arenas,Ikot}.

Thus, it was clear that the study of systems with position-dependent mass, as well as information theory, is of great interest in physics. However, few studies have been carried out involving both subjects. Therefore, in this work, we study Fisher information and Shannon entropy for a mass system position-dependent with a hyperbolic barrier potential.

The work is organized as follows: In section \ref{sec1}, we show some fundamental concepts of the one-dimensional PDM problem ordered by BenDaniel-Duke. In addition, we found the analytical solutions of the Schrödinger-type equation for a solitonic mass distribution when subjected to a barrier potential $V(x)=V_1 \coth^2(x)+V_2\mathrm{csch}^2(x)$. In section \ref{sec2}, we present the basic concepts of Shannon entropy and Fisher information and apply them to our confined particle system.  Finally, in section \ref{sec3}, we make the final remarks and discuss our results.

\section{Solution of Schrödinger  problem for a position-dependent mass}
\label{sec1}

To solve a problem in which position-dependent mass, it is not enough to simply substitute the mass profile in the Schrödinger equation, as a problem arises regarding the hermiticity of the kinetic energy operator $\hat{T}$. The best way to solve this problem is to use a symmetric operator. Currently, there are several types of kinetic energy requests that are based on a symmetric operator that guarantees that the Hamiltonian operator is Hermitian. The five ordering most used are BenDaniel-Duke \cite{BenDaniel}, Gora-Willian \cite{GoraW}, Zhu-Kroemer  \cite{ZhuKroemer}, Li-Kuhn \cite{LiKuhn} and the most current, Mustafa-Mazharimousavi \cite{Mustafa}. It is interesting to note in 1983, Von Roos \cite{Roos}, proposes a generalization of the symmetrical kinetic energy operator in the form
\begin{eqnarray}\label{1}
\hat{T}=\frac{\hbar}{4}\bigg[m^{\alpha}(\vec{r}) \hat{p}m^{\beta}(\vec{r})\hat{p}m^{\gamma}(\vec{r})+m^{\gamma}(\vec{r})\hat{p}m^{\beta}(\vec{r})\hat{p}m^{\alpha}(\vec{r})\bigg],
\end{eqnarray}
where $\alpha $, $ \beta$, $ \gamma $ are constants and are named Von Roos ordering or ambiguity parameters. These parameters must satisfy the relation $ \alpha+\beta+\gamma=-1 $ \cite{Roos}.

Currently, the ordering of kinetic energy that draws the most attention is that of BenDaniel-Duke, for its simplicity and significant results \cite{Cunha,Falaye,Sun2015,Dong2016,Navarro}. The kinetic energy operator proposed by BenDaniel-Duke is \cite{BenDaniel}
\begin{eqnarray}
\hat{T}=\frac{1}{2m(x)}\hat{p}^{2}+\frac{i\hbar }{2}\frac{m'(x)}{m^2(x)}\hat{p},
\end{eqnarray}
where $m'(x)=dm(x)/dx$. Note that we can get to equation (\ref{1}) by doing $\alpha=\gamma=0$ and $\beta=-1$, obeying the relation of the Von Roos parameters \cite{Roos}. 

Then, with the BenDaniel-Duke ordering for kinetic energy, we have the Schrödinger-type equation for a particle with position-dependent mass, subject to any potential $V(x)$ \cite{Cunha,BenDaniel}
\begin{eqnarray}\label{2}
\Big[\frac{1}{2m(x)}\hat{p}^{2}+\frac{i\hbar }{2}\frac{m'(x)}{m^2(x)}\hat{p}+V(x)\Big]\psi(x)=E\psi(x).
\end{eqnarray}
where $\psi(x)$ is the standing wave function and $V(x)$ is the potential that defines the system.

\subsection{Solitonic mass profile}

\begin{figure}
\begin{center}
\begin{tabular}{ccc}
\includegraphics[height=6cm]{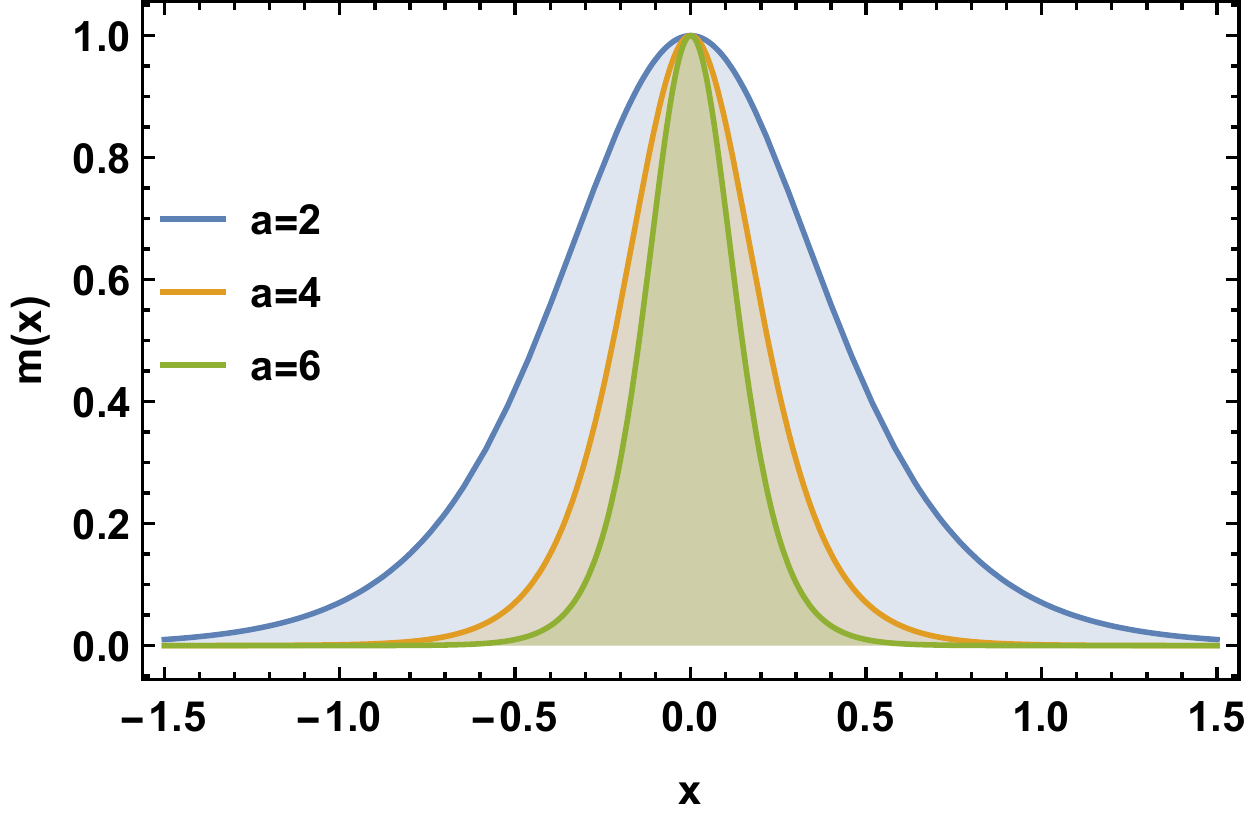}
\end{tabular}
\end{center}
\caption{Behavior of solitonic mass distribution with $m_{0}$ constant.
\label{fig1}}
\end{figure}

The profile of the mass function $m(x)$ defines the change in the particle mass with the variation of position, that is, this $m(x)$ mass profile is very important for the definition of the equation that will describe the motion of the particle. We propose a solitonic mass profile, which is described in the form \cite{Cunha}
\begin{equation}\label{mass}
m(x)=m_{0}\mathrm{sech}^{2}(ax).
\end{equation}
where $m_{0}$ is the asymptotic mass of the system when $x\rightarrow0$ and $a$ 
is the parameter that controls the width of the mass distribution. This mass profile is interesting, because when $x\rightarrow\infty$ our mass distribution is $m(x)\rightarrow0$, as seen in figure \ref{fig1}.
 
\begin{figure}
\begin{center}
\begin{tabular}{ccc}
\includegraphics[height=6cm]{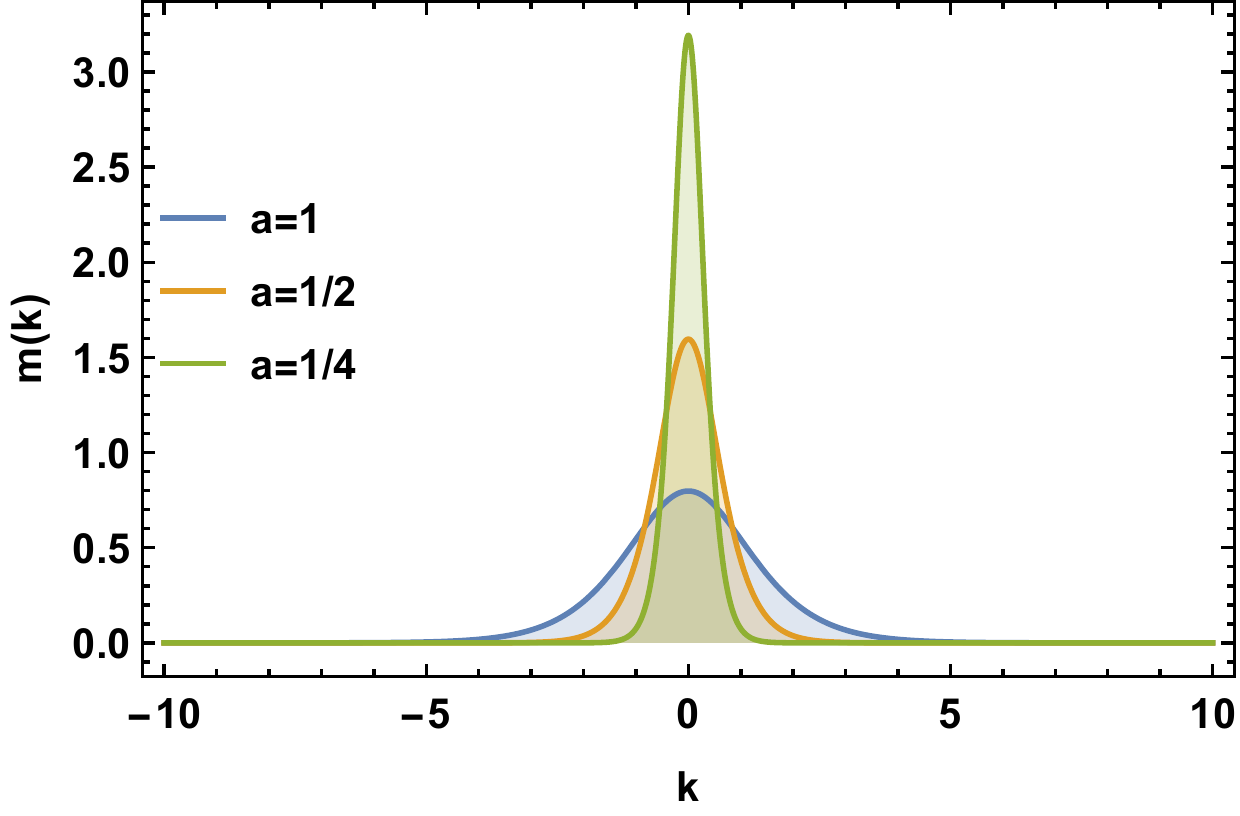} 
\end{tabular}
\end{center}
\caption{Behavior of mass distribution in space-$k$ with $m_{0}$ constant.
\label{fig2}}
\end{figure}

We make this choice because it is an appropriate representative of a solitonic distribution (soliton-like mass) found in various models of condensed matter and low-energy nuclear energy physical \cite{Cunha,Bagchi}. Solitons are structures that arise in a non-linear theory. These structures are interesting because they have finite energy and keep their form unchanged when interacting with another soliton \cite{Heeger,Kartashov,Rajaraman}. 
The mass distribution $m(x)$ can also be represented in k-space \cite{Lima2021}, in the form
\begin{eqnarray}
m(k)=\sqrt{\frac{\pi}{2}}\frac{m_{0}k}{a^{2}}\mathrm{csch}\Big(\frac{k\pi}{2a}\Big).
\end{eqnarray}
In Fig.\ref{fig2} we give the behavior of the mass distribution in the reciprocal space.

The mass profile representing in space-$k$ gives us the energy of the dispersion \cite{Lima2021},
\begin{eqnarray}
E(k)=\sqrt{\frac{2}{\pi}}\frac{a^{2}\hbar^{2}}{m_{0}}\Big[{k^{2}} _{2}F_{1}\Big(\frac{1}{2}, \frac{3}{2};  \frac{3}{2}; \frac{k^{2}}{4} \Big)-\cosh(k)\Big].
\end{eqnarray}

Replacing Eq.(\ref{mass}) in Eq.(\ref{2}), we have \cite{Cunha}
\begin{eqnarray}\label{3}
\Big\{\frac{d^2}{dx^2}+2 \tanh(x)\frac{d}{dx}+\frac{2m_{0}}{a^2\hbar^2}[E-V(x)]\mathrm{sech}^2(x)\Big\}\psi(x)=0,
\end{eqnarray}
where is made $ax\rightarrow x$, for simplicity. With equation (\ref{3}), it is possible to analyze the behavior of a one-dimensional particle, whose mass depends on the position.

\subsection{Solution to a hyperbolic barrier potential}

\begin{figure}
\begin{center}
\begin{tabular}{ccc}
\includegraphics[height=6cm]{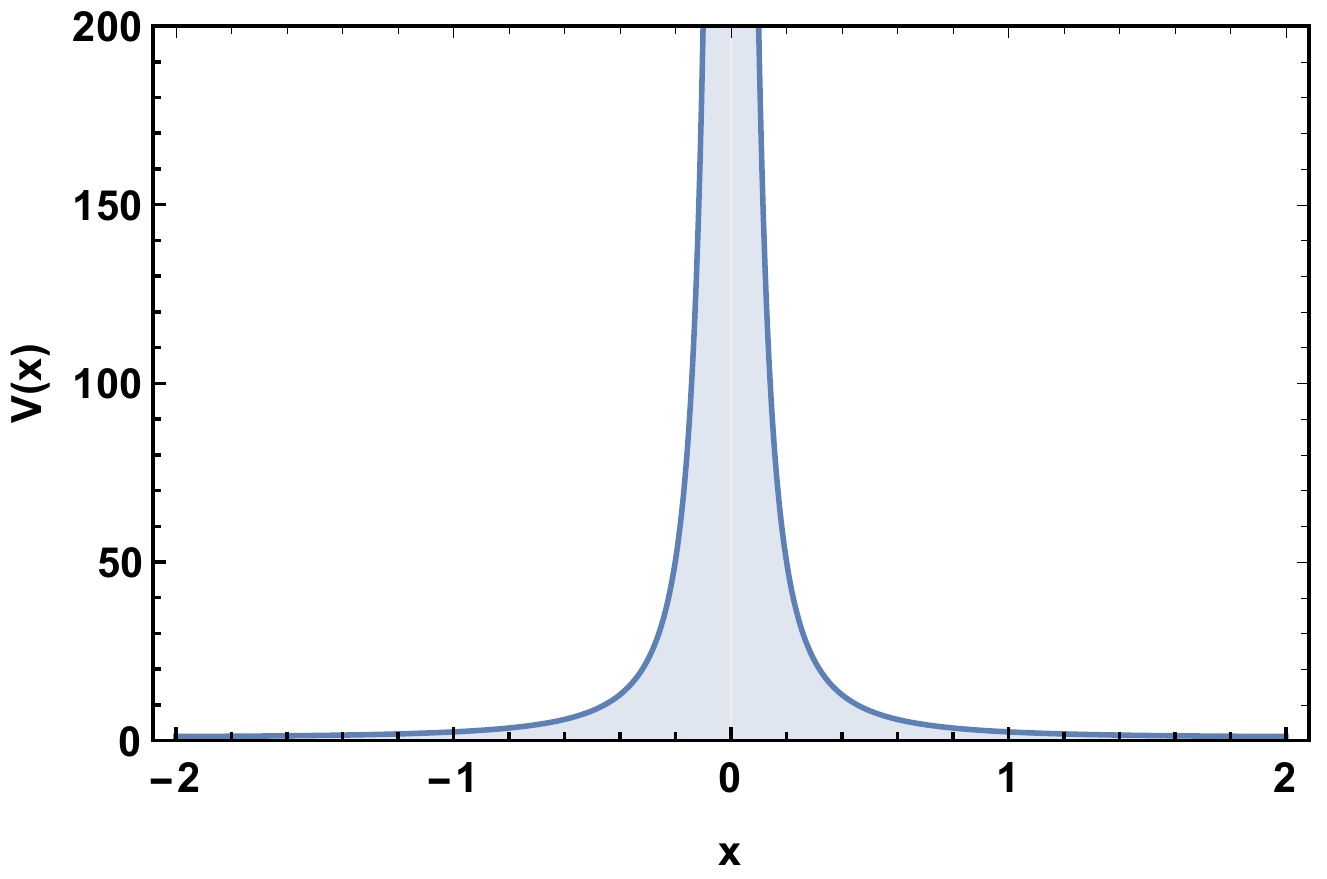} 
\end{tabular}
\end{center}
\caption{Representation of the hyperbolic potential $V(x)$ for the positive $V_1$ and $V_2$ values.
\label{fig3}}
\end{figure}

As our interest is to study the behavior of a quantum-mechanical system with PDM in a barrier potential, we consider the hyperbolic potential as follows
\begin{eqnarray}\label{4}
V(x)=V_1\coth^2(x)+V_2 \mathrm{csch}^2(x),
\end{eqnarray}
where $V_1$ and $V_2$ are parameters that define a barrier potential. We plot the behavior of the potential (\ref{4}) in the figure \ref{fig3}.

Even with a null potential $V(x)=0$ (which identifies a free particle), the particle is not entirely free, because as we saw in the previous section, the mass profile (\ref{mass}) acts as a kind of confining potential \cite{Cunha}. Thus, with the potential (\ref{4}), we guarantee a new interaction of the particle that will not be confined to the origin.

To find the solutions of the Schrödinger equation (\ref{3}), we do 
\begin{equation}
\psi(x)=\cosh^v(x)\phi(x),
\end{equation}
where $v$ is an arbitrary parameter, to give us
\begin{eqnarray}\label{5}
\frac{d^2}{dx^2}\phi(x)+2(v+1)\tanh(x)\frac{d}{dx}\phi(x)+ \Big\{v(v+2)\tanh^2(x)+\Big[v+ \frac{2m_0}{a^2\hbar^2}[E-V(x)]\Big]\mathrm{sech}^2(x)\Big\}\phi(x)=0.
\end{eqnarray}

For simplicity, we make the transformation $x \rightarrow z$, in the form
\begin{equation}
\cos(z)=\mathrm{sech}(x),
\end{equation} 
where $x \in (-\infty,\infty) \rightarrow z \in (-\pi/2,\pi/2)$. Thus, we obtain Eq.(\ref{5}) in terms of the variable $z$,
\begin{eqnarray}
\frac{d^2}{dz^2}\phi(z)+(1+2v)\tan(z)\frac{d}{dz}\phi(z)+\Big\{v+v(v+2)\tan^2(z)+\frac{2m_0}{a^2\hbar^2}[E-V(z)]\Big\}\psi(z)=0.
\end{eqnarray}

Choosing $v=-1/2$ eliminates the first derivative in $\phi$, which results in \cite{Cunha}
\begin{eqnarray}\label{6}
-\frac{d^2}{dz^2}\phi(z)+\Big[\frac{1}{2}+\frac{3}{4}\tan^2(z)+\tilde{V}(z)\Big]\phi(z)=\varepsilon\phi(z),
\end{eqnarray}
where we define
\begin{eqnarray}
\tilde{V}(z)=\frac{2m_0}{a^2\hbar^2}V(z)\qquad \mathrm{and}  \qquad \varepsilon=\frac{2m_0}{a^2\hbar^2}E.
\end{eqnarray}

\begin{figure}
\begin{center}
\begin{tabular}{ccc}
\includegraphics[height=6cm]{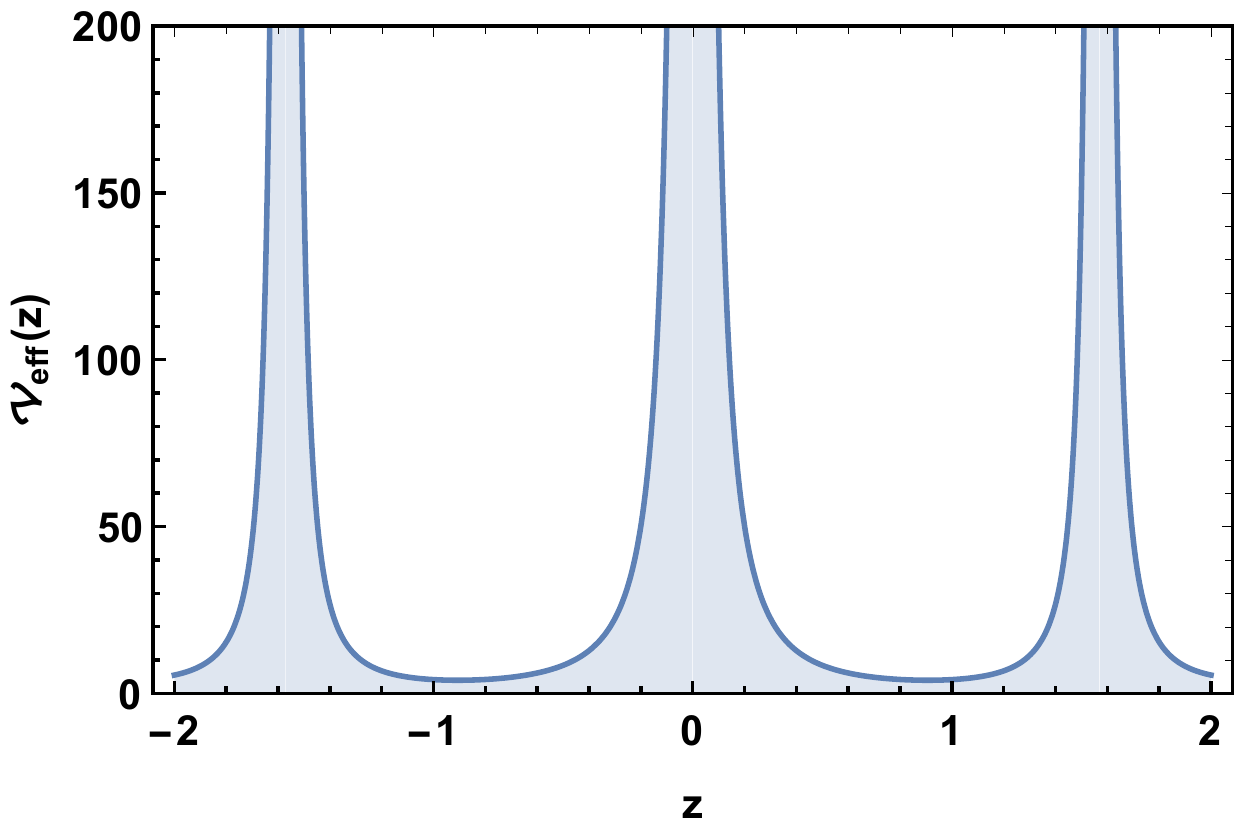} 
\end{tabular}
\end{center}
\caption{Representation of effective potential $\mathcal{V}_{eff}(z)$ for the positive $V_1$ and $V_2$ values.
\label{fig4}}
\end{figure}

The Eq.(\ref{6}) allows us to find symmetric and antisymmetric solutions, as long as $\tilde{V}(z)$, i. e., $V(x)$ is symmetric \cite{Cunha}. Therefore, equation (\ref{6}) is equivalent to a regular stationary equation of the Schrodinger type, with a constant mass $m_0$, and an effective confining potential of the type
\begin{eqnarray}\label{7}
\mathcal{V}_{eff}(z)={\frac{1}{2}}+{\frac{3}{4}}\tan^2(z)+\tilde{V}(z),
\end{eqnarray}
where the dynamics is restricted to the interval $z=(-\pi/2,\pi/2)$, subject to the boundary conditions $\phi(z=\pm\pi/2)=0$. As we can see in Eq.(\ref{7}), even though the potential is null, we will still have an effective potential confining the particle. For our hyperbolic potential (\ref{4}), we have $V(z)= (V_1 +V_2)\mathrm{csc}^2 (z)-V_2$, and by equation (\ref{7}) the effective potential is
\begin{eqnarray}
\mathcal{V}_{eff}(z)={\frac{1}{2}}+{\frac{3}{4}}\tan^2(z)+ \frac{2m_0}{a^2\hbar^2}\bigg[ (V_1 +V_2)\mathrm{csc}^2 (z)-V_2 \bigg],
\end{eqnarray}
which is represented in figure \ref{fig4}.

Because the potential is discontinuous in origin, the solutions of Eq.(\ref{6})  are extremely difficult to find. We can write the solutions in terms of $\psi(x)$, which are
\begin{eqnarray}\label{8}
\psi^1 (x)&=& C_1\Big[\frac{\mathrm{coth}(a x)^{-\frac{1}{2}(1+\varsigma)}}{\sinh^2(a x)}\Big]
{_2F}_1\Big(1+\frac{1}{4}(\vartheta-\varsigma), 1-\frac{1}{4}(\vartheta+\varsigma),1-\frac{1}{2}\varsigma;\mathrm{coth}^2(a x)\Big),\\
\psi^2 (x)&=& C_2\Big[\frac{\mathrm{coth}(a x)^{-\frac{1}{2}(1-\varsigma)}}{\sinh^2(a x)}\Big]
{_2F}_1\Big(1+\frac{1}{4}(\vartheta+\varsigma), 1-\frac{1}{4}(\vartheta-\varsigma),1+\frac{1}{2}\varsigma;\mathrm{coth}^2(a x)\Big),
\end{eqnarray}
where 
\begin{eqnarray}
\vartheta&=&\sqrt{ 4(V_1 +V_2) +1},\nonumber\\
\varsigma&=& \sqrt{1+4\Big[V_1+V_2-\Big(\frac{2n\varrho-4n^2+2\varrho-8n-4}{\kappa^2}\Big)\Big]},\nonumber\\
\varrho&=&\sqrt{4\kappa^2(V_1+V_2)+1},
\end{eqnarray}
with $\kappa^2=2m_o/a^2\hbar^2$. Only one solution is acceptable $\psi^1(x)$, because, for the wave functions to be physically acceptable, the normalization of these functions must be preserved. In this way, the wave function $\psi^2(x)$ tends to diverge at the origin of the system. With equation (\ref{8}) it is noted that energy is quantized, with energy levels
\begin{eqnarray}
E_n=\Big[\frac{2\varrho(n-1)-4(n^2+2n+1)}{\kappa^2}\Big]-V_1,\qquad \mathrm{with}  \qquad n=0,1,2,...
\end{eqnarray}

For a better analysis of equation (\ref{8}), we define $V_1=V_2=\kappa=1$, since they are positive constants, reducing Eq.(\ref{8}) to
\begin{eqnarray}
\psi^1_n= C^n_1\Big[\frac{\tanh(a x)^{2n+1}}{\sinh^2(a x)}\Big]
{_2F}_1\Big(\frac{3}{2}-n, -n,\frac{1}{2}-2n;\mathrm{coth}^2(a x)\Big),
\end{eqnarray}
with $C^n_1$ to be determined by system normalization. 

The solutions for the three lowest energy states found normalized, are 
\begin{eqnarray}\label{solu}
\psi_0^1 (x)&=&\sqrt{\frac{35a}{4}}\tanh^2(a x) \mathrm{sech}^{2}(ax),\nonumber\\
\psi_1^1 (x)&=&\sqrt{\frac{6237a}{32}}\Big[\frac{4\cosh^2(ax)}{9}-1\Big]\tanh^2(a x) \mathrm{sech}^{4}(ax),\nonumber\\
\psi_2^1 (x)&=&\sqrt{\frac{920205a}{256}}\Big[\frac{24\cosh^4(ax)-132\cosh^2(ax)}{143}+1\Big]\tanh^2(a x) \mathrm{sech}^{6}(ax).
\end{eqnarray}

\begin{figure}
\begin{center}
\begin{tabular}{ccc}
\includegraphics[height=5cm]{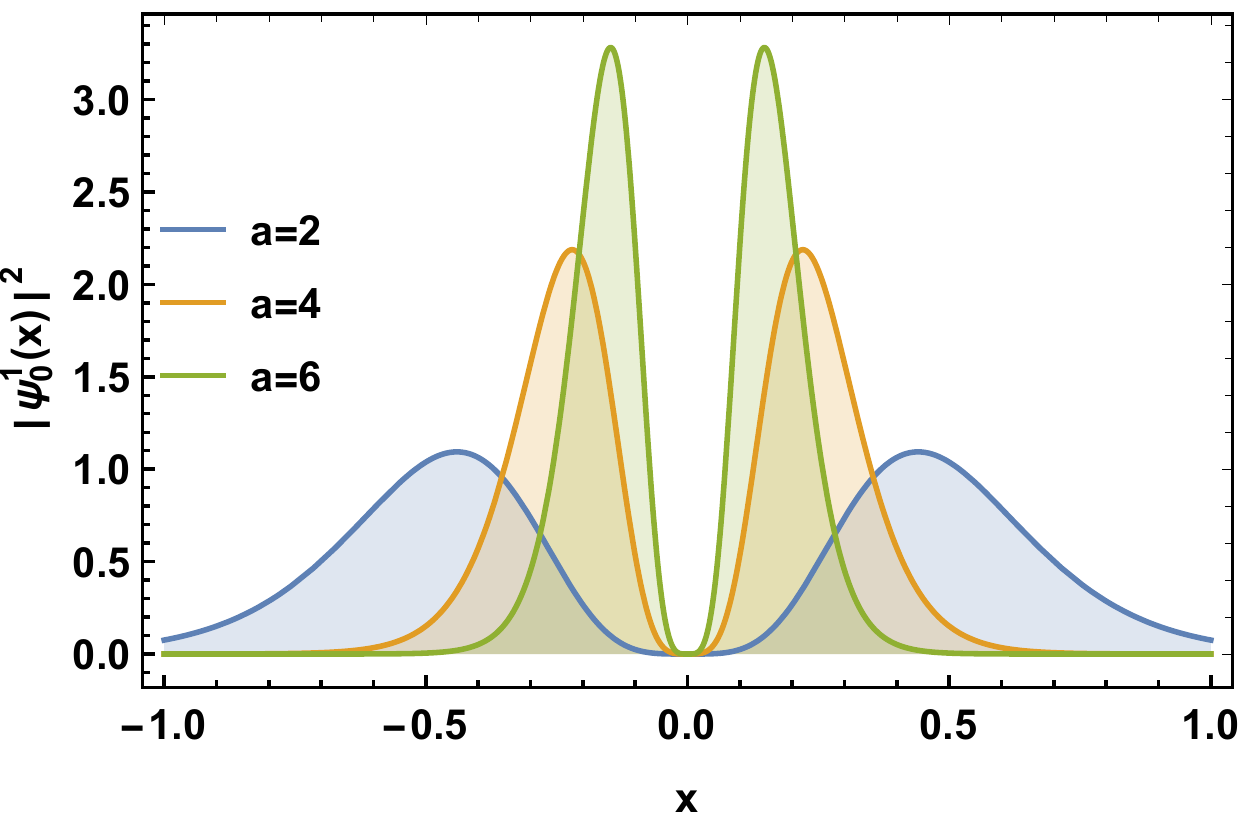} \\
(a) \\
\includegraphics[height=5cm]{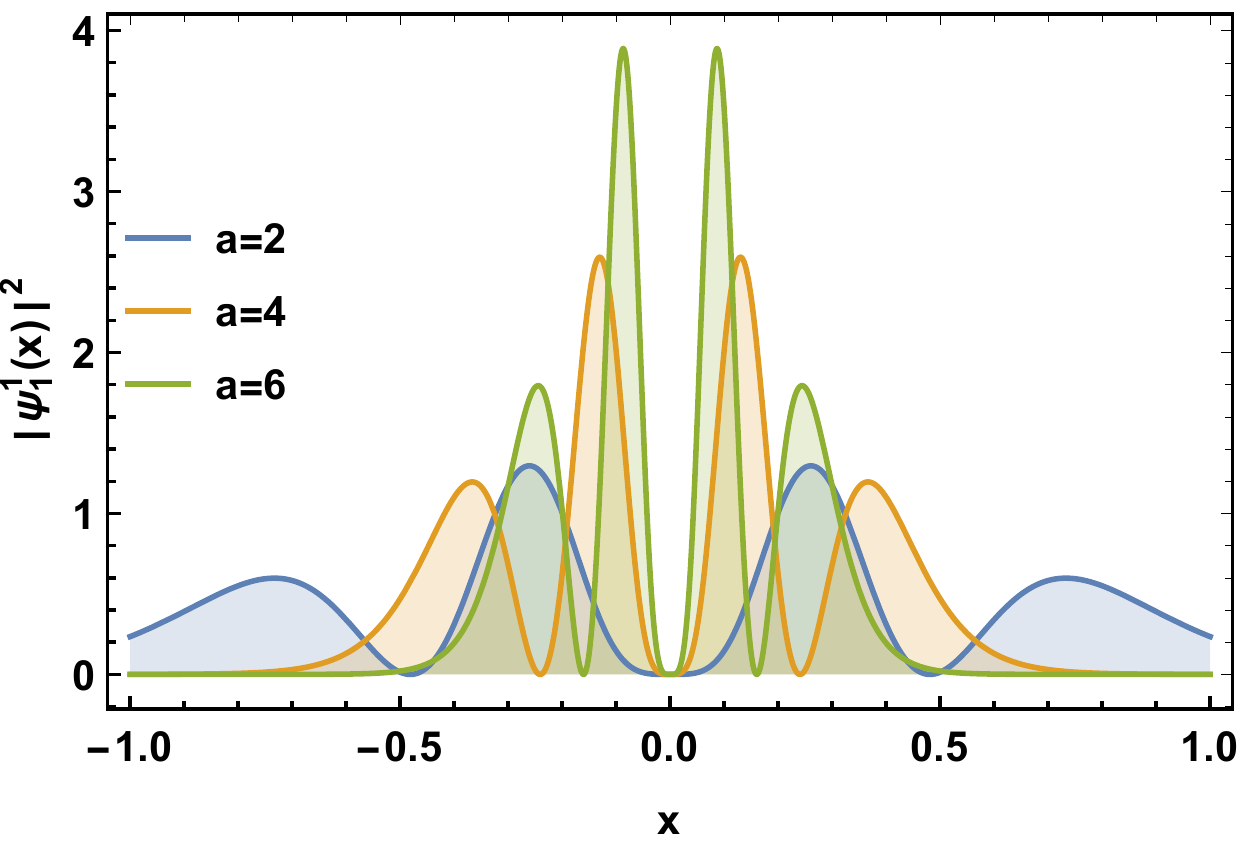} 
\includegraphics[height=5cm]{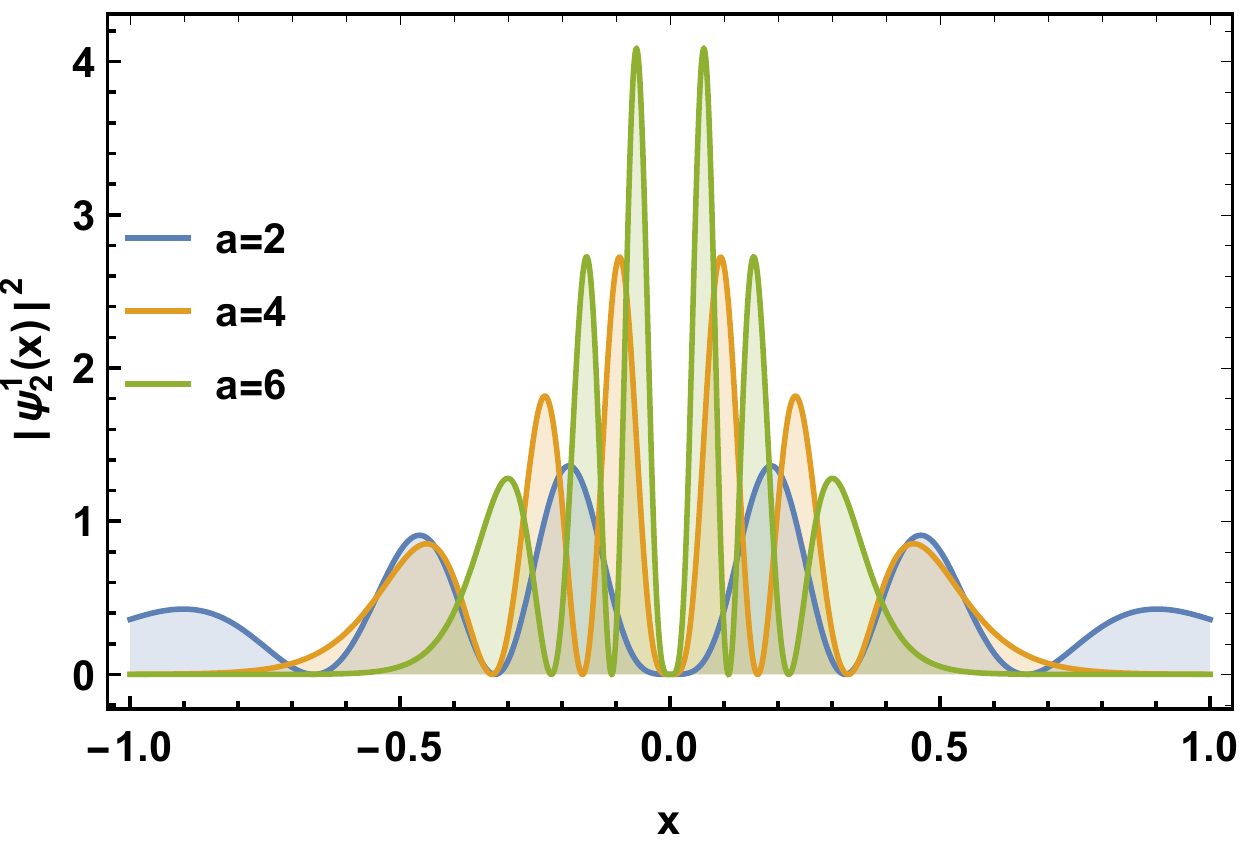}\\
(b) \hspace{6 cm}(c)
\end{tabular}
\end{center}
\caption{Behavior of probability densities. (a) $n=0$. (b) $n=1$. (c) $n=2$.
\label{fig5}}
\end{figure}

The behavior of these wave functions is observed by analyzing their probability densities $\rho(x)=\vert\psi(x)\vert^2$, as represented in the figure \ref{fig5}.

\section{Information Theories}
\label{sec2}

With the advancement of communication technology, there was an increase in the studies of theories that involve the transmission of information \cite{Nalewajski,Nagaoka,Wang,Lian,Falaye0}. The interest in the study of these theories comes from the fact that they can be applied to quantum systems \cite{Rothstein,Zou}, some of these studies are linked to modern quantum communication \cite{Falaye,Serrano}, computing and the Density Functional Theory known by its acronyms \cite{Burke}. The main theories related to information transmission that can be applied to a quantum system are Shannon entropy \cite{Gadre1985} and Fisher information \cite{Frieden1}.

\subsection{Shannon's Theory}

The initial concept of entropy in Thermodynamics refers to a measure of the irreversibility of the physical system \cite{Callen}, or a measure associated with the degree of disorder of the system \cite{Callen,Pathria}. Following this same line of reasoning Claude E. Shannon in 1948, in the work entitled \textit{``A mathematical theory of communication"} \cite{Shannon}, describes for the first time entropy as an element of the theory of information and communication. Shannon entropy is a quantity that measures the uncertainty in a given probability distribution.

The interesting thing about this theory is that it is related to the probability density of an information system, defined as entropic densities \cite{Shannon}, which can be related to the probability density found in quantum systems $\vert\psi( x)\vert^{2}$ \cite{Navarro,Born,Hirschmann,Beckner}. With these definitions, the entropic densities with respect to the position $\rho^n_s(x)$ and the momentum $\rho^n_s(p)$ can be represented \cite{Shannon}
\begin{equation}
{\rho^n_s(x)=|\psi(x)|^2 \ln|\psi(x)|^2},
\end{equation}
\begin{equation}
{\rho^n_s(p)=|\phi(p)|^2 \ln|\phi(p)|^2}.
\end{equation} 

Thus, for a probability density of a continuous system, the Shannon entropy can be defined for the position space $S^n_{x}$ and for the momentum space $S^n_{p}$ \cite{Shannon}
\begin{equation}
S^n_{x}=-\int_{-\infty}^{\infty}\vert\psi_{n}(x)\vert^{2}\ln\vert\psi_{n}(x)\vert^{2}dx,
\end{equation} 
\begin{equation}
S^n_{p}=-\int_{-\infty}^{\infty}\vert\Phi_{n}(p)\vert^{2}\ln\vert\Phi_{n}(p)\vert^{2}dp.
\end{equation}

Beckner, Bialynicki-Birula, and Mycielski in 1975 obtained the relation of entropic uncertainty related to position and momentum, which became known as BBM uncertainty \cite{Beckner,Bialy}. The uncertainty relation of BBM is given by \cite{Beckner,Bialy} 
\begin{equation}
S^n_{x}+S^n_{p}\geq D(1+\ln\pi),
\end{equation}
where $D$ represents the spatial dimension of the system.

\subsubsection{Shannon's Entropy Analysis}

\begin{figure}
\begin{center}
\begin{tabular}{ccc}
\includegraphics[height=5cm]{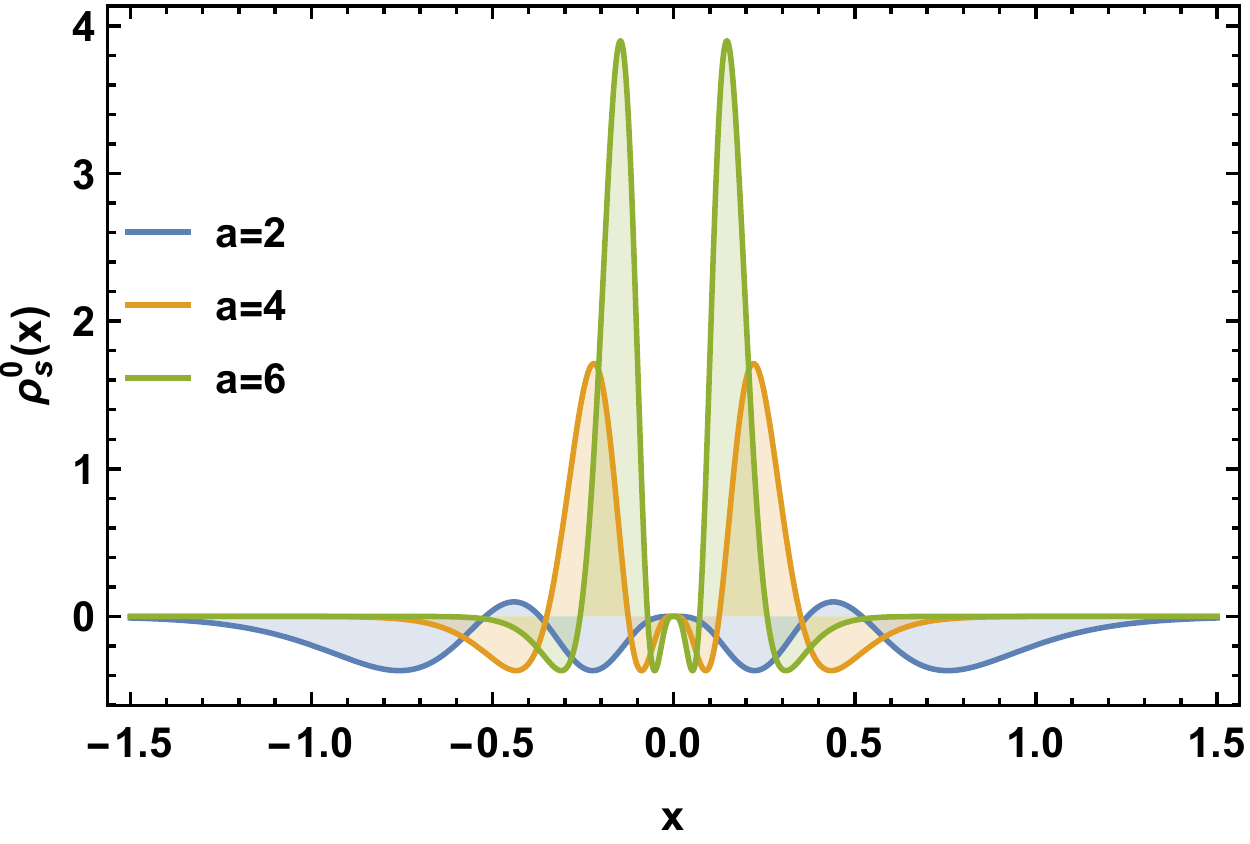} \\
(a) \\
\includegraphics[height=5cm]{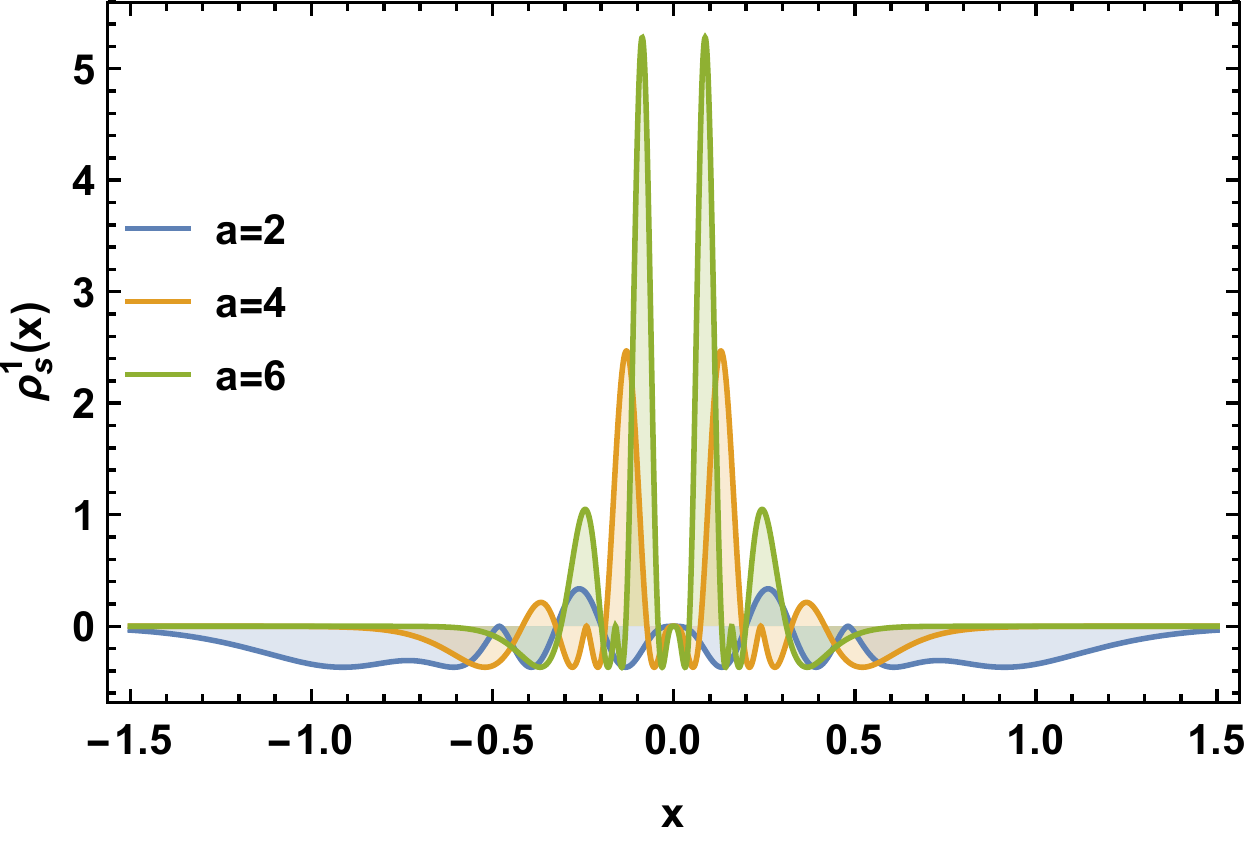} 
\includegraphics[height=5cm]{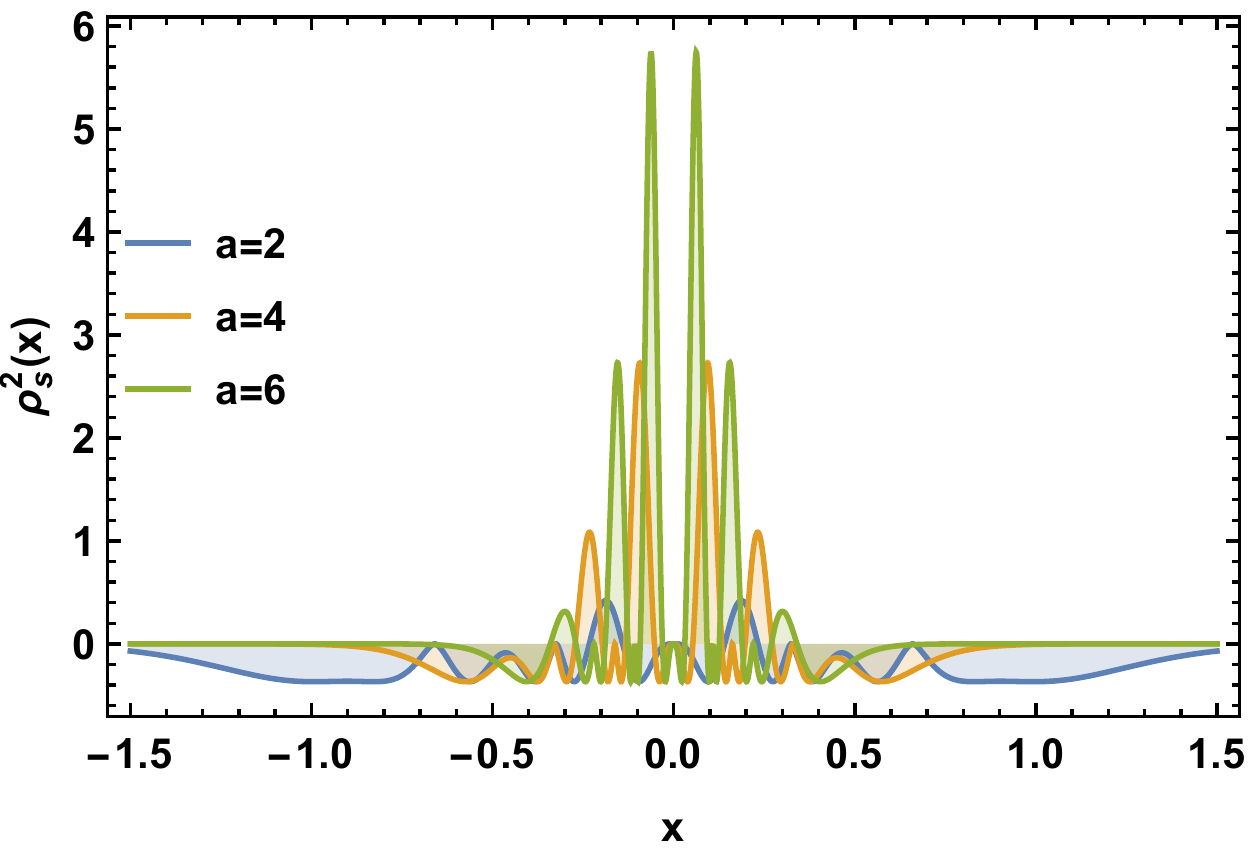}\\
(b) \hspace{6 cm}(c)
\end{tabular}
\end{center}
\caption{Behavior of the entropy densities $\rho_s(x)$. (a) $n = 0$. (b) $n = 1$. (c) $n = 2$.
\label{fig6}}
\end{figure}

We apply Shannon's Entropy to analyze solutions to our one-particle problem
with position-dependent mass in a hyperbolic potential. To start with, it is interesting to observe the entropy behavior of the wave functions (\ref{solu}), we do this by analyzing their entropic densities $\rho_s(x)=|\psi(x)|^2 \ln|\psi(x)|^2 $, represented in the figure \ref{fig6}.

With the help of the Fourier Transform, we obtain the eigenfunctions for the momentum space, in the three lowest energy states of the system
\begin{eqnarray}\label{solup}
\phi_0^1 (p)&=&\sqrt{\frac{35a\pi}{8}}\frac{p}{6a^4}(-2a^2+p^2)\mathrm{csch}\Big(\frac{p\pi}{2a}\Big),\nonumber\\
\phi_1^1 (p)&=&\sqrt{\frac{156237a\pi}{64}}\frac{p}{1080a^6} (16a^4-80a^2p^2+9p^4) \mathrm{csch}\Big(\frac{p\pi}{2a}\Big),\nonumber\\
\phi_2^1 (p)&=&\sqrt{\frac{920205a\pi}{412}}\frac{p}{720720a^8} (6528a^6-12152a^4p^2+3542a^2p^4-143p^6) \mathrm{csch}\Big(\frac{p\pi}{2a}\Big).
\end{eqnarray}

\begin{figure}
\begin{center}
\begin{tabular}{ccc}
\includegraphics[height=5cm]{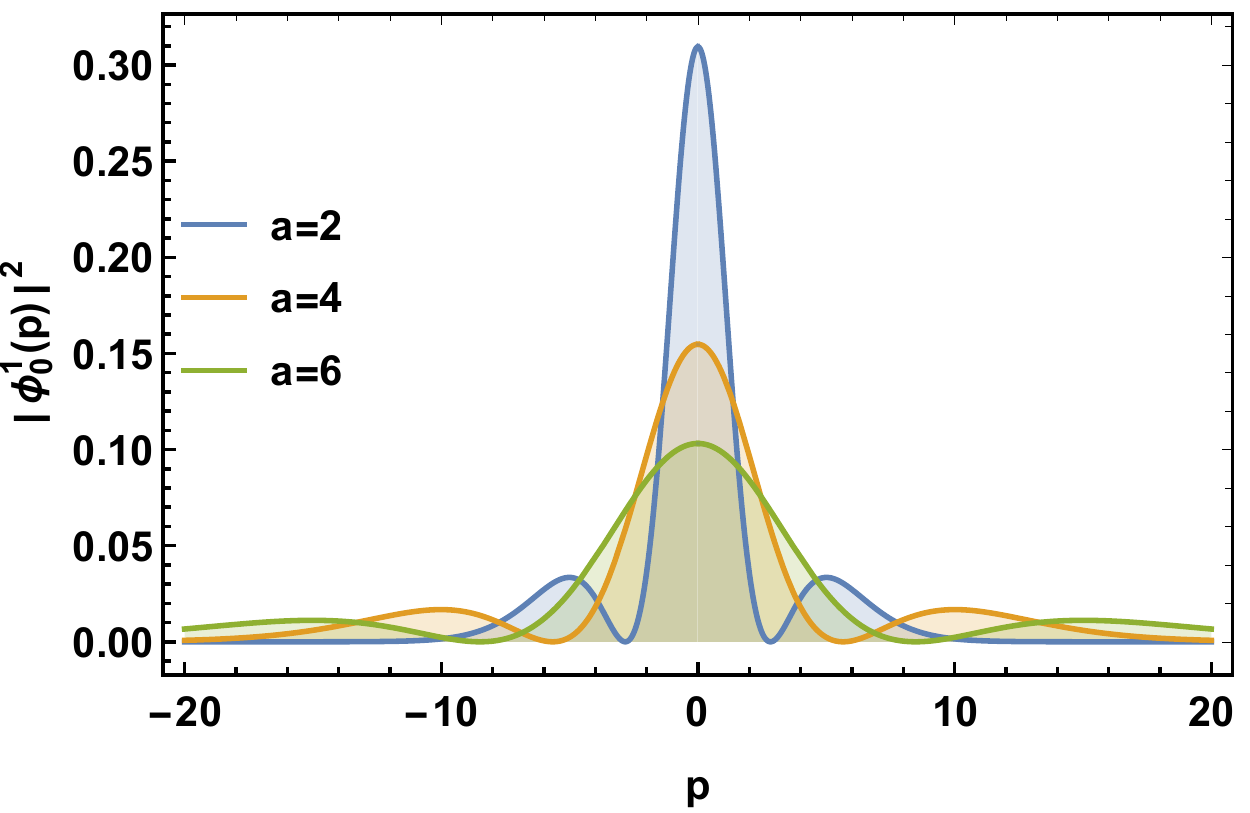} \\
(a) \\
\includegraphics[height=5cm]{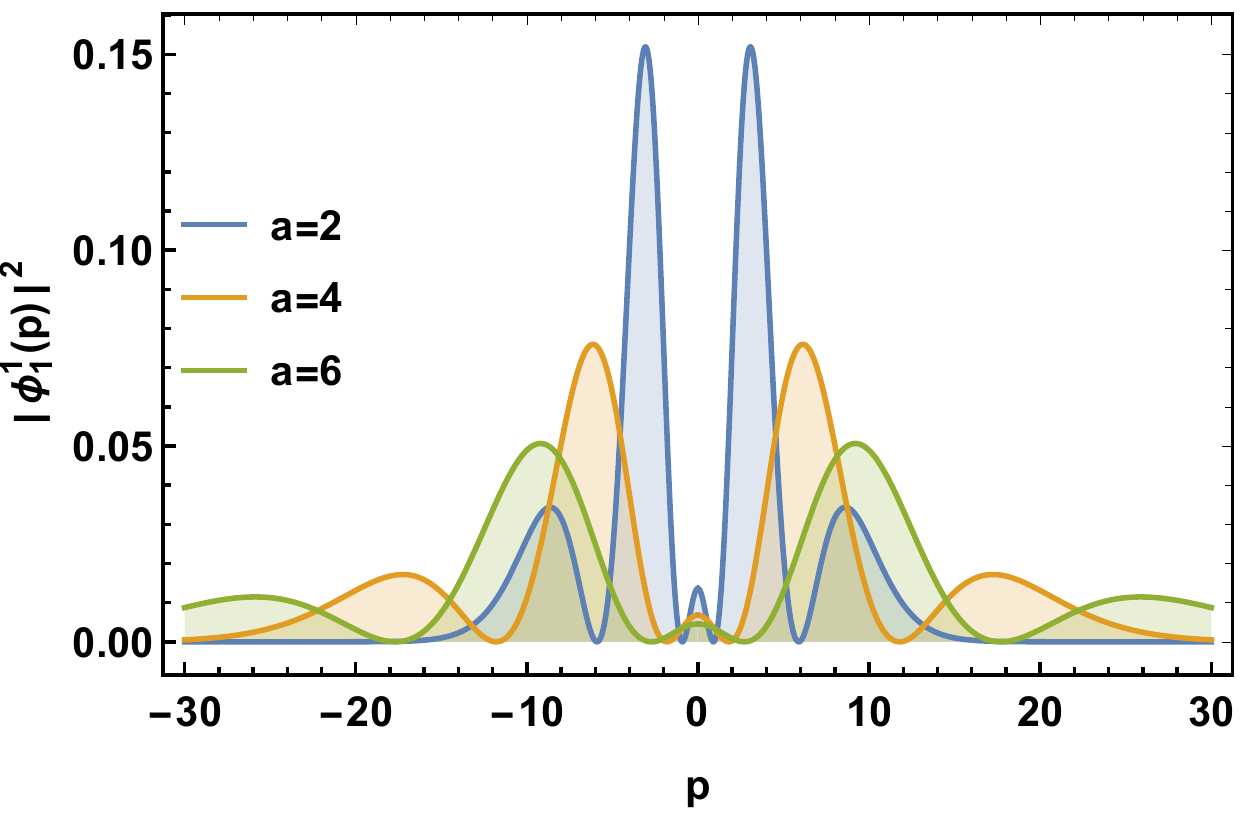} 
\includegraphics[height=5cm]{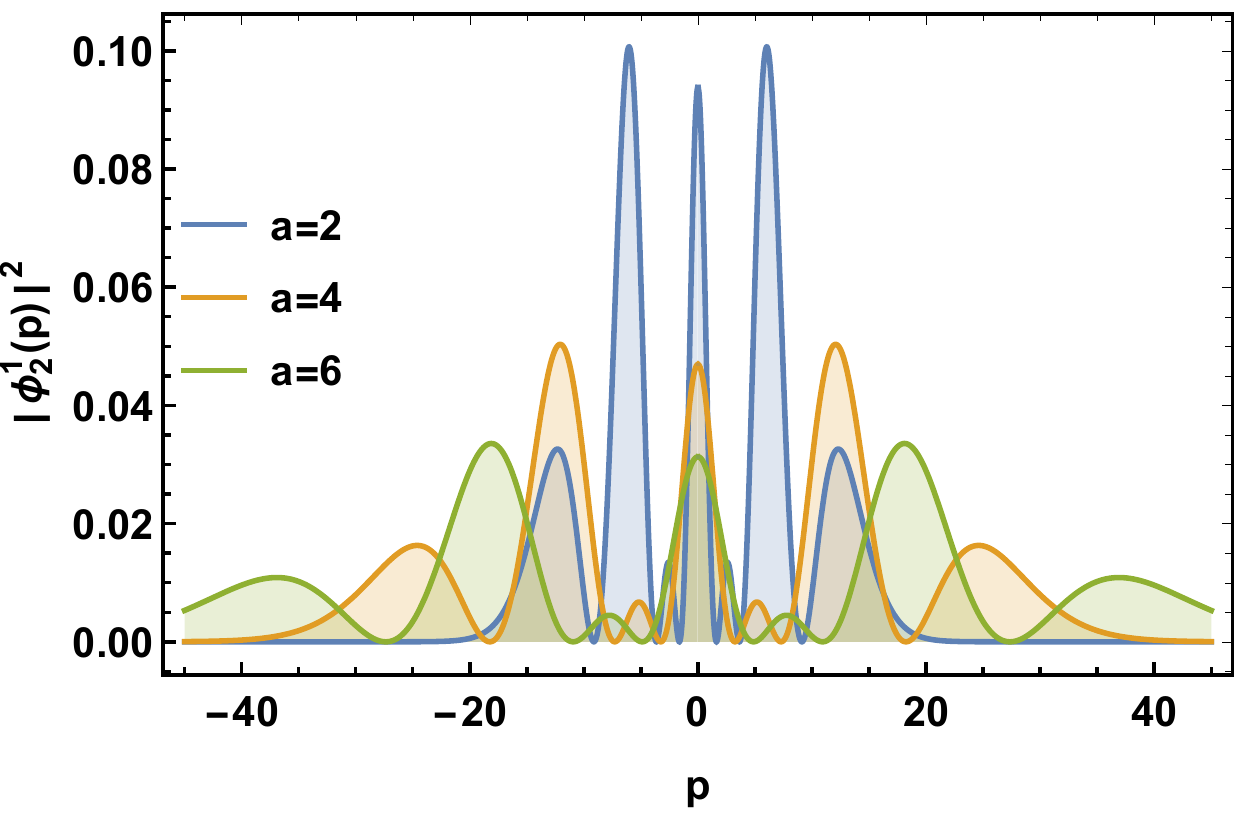}\\
(b) \hspace{6 cm}(c)
\end{tabular}
\end{center}
\caption{Behavior of probability densities in momentum space. (a) $n = 0$. (b) $n = 1$. (c) $n = 2$.
\label{fig7}}
\end{figure}

\begin{figure}
\begin{center}
\begin{tabular}{ccc}
\includegraphics[height=5cm]{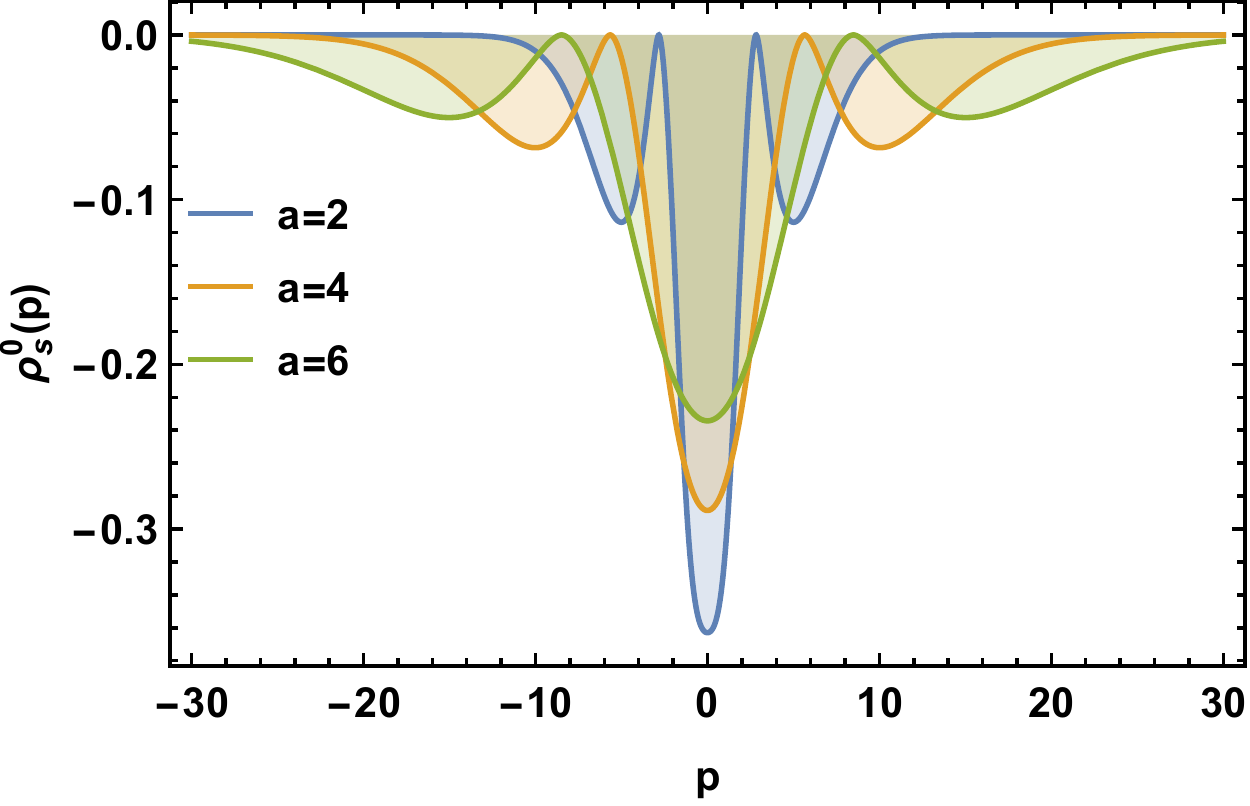} \\
(a) \\
\includegraphics[height=5cm]{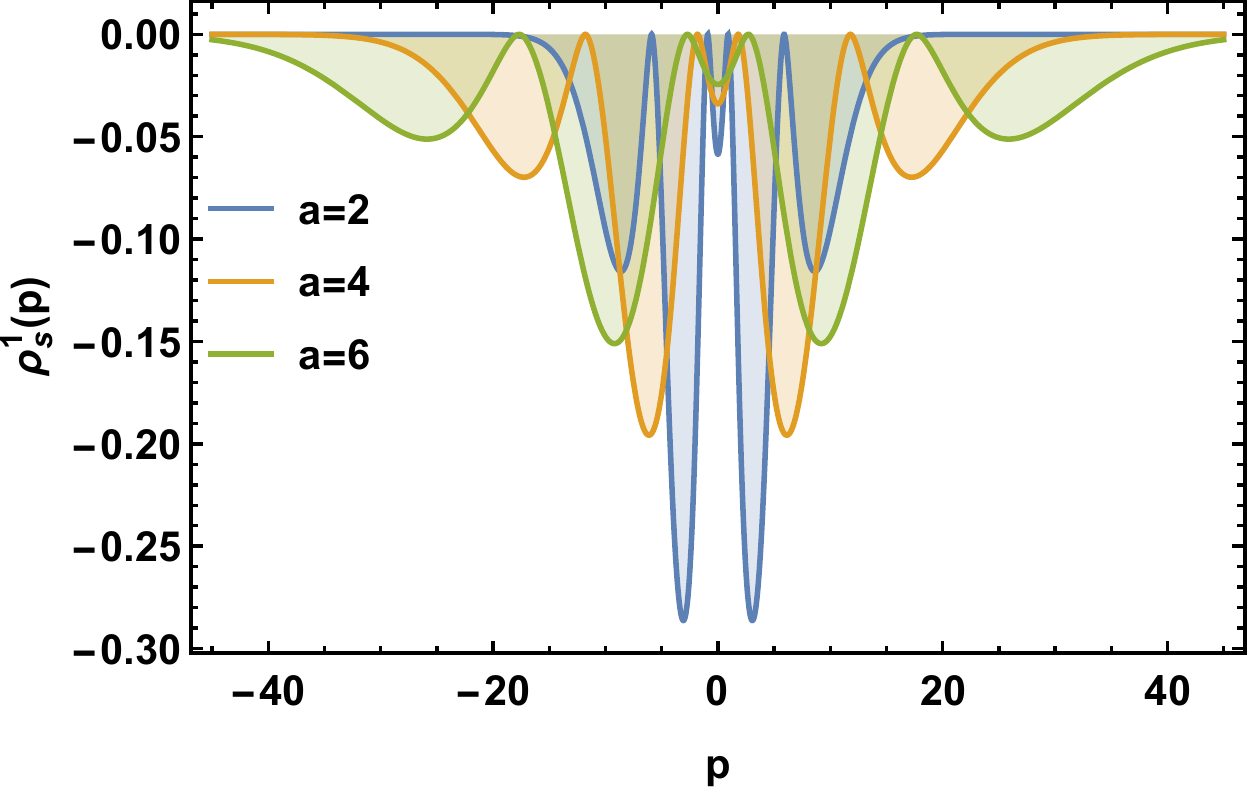} 
\includegraphics[height=5cm]{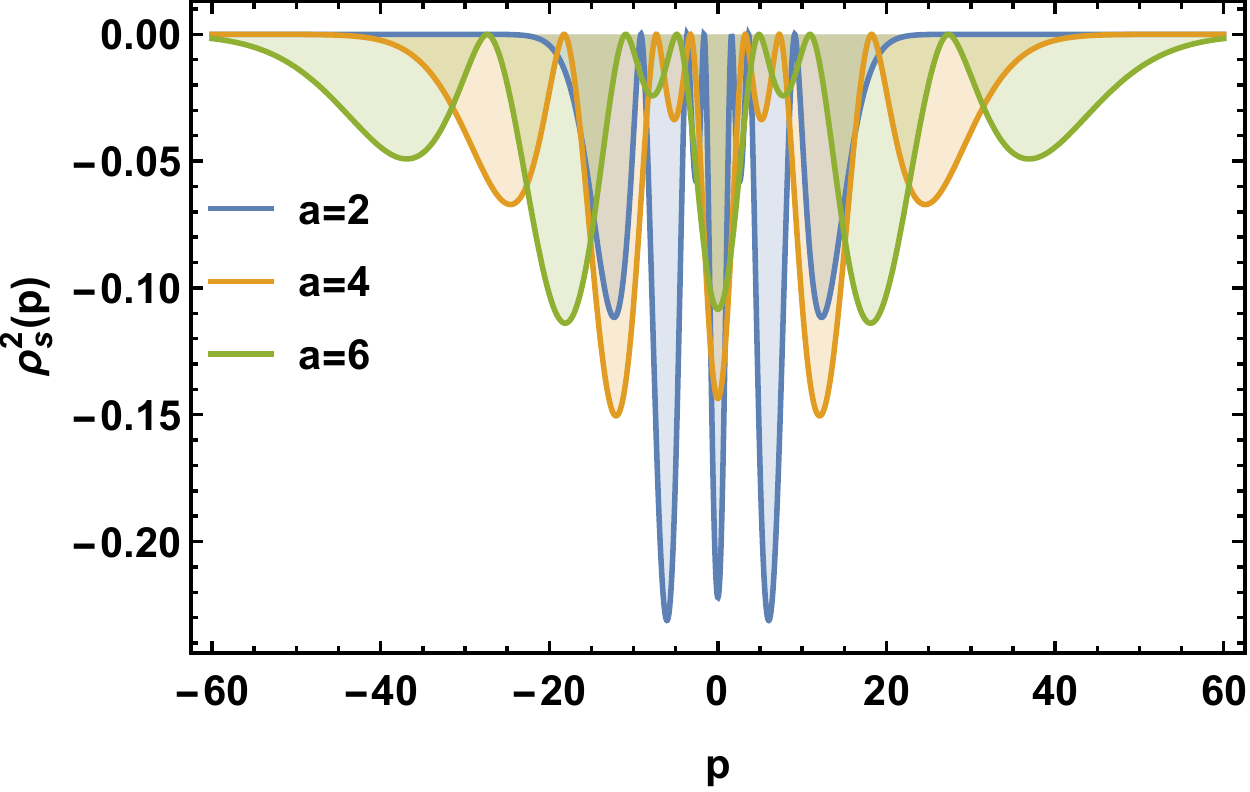}\\
(b) \hspace{6 cm}(c)
\end{tabular}
\end{center}
\caption{Behavior of the entropy densities in the momentum space $\rho_s(p)$. (a) $n = 0$. (b) $n = 1$. (c) $n = 2$.
\label{fig8}}
\end{figure}

In the Fig.\ref{fig7} we plot the probability densities in the momentum space $\rho(p)=|\phi(p)|^2$. The entropic densities $\rho_s(p)=|\phi(p)|^2 \ln|\phi(p)|^2$, are analyzed through the figure \ref{fig8}. 
With the table \ref{tab1}, the numerical study of Shannon's entropy was carried out considering the eigenfunctions in the space of position and momentum.

\begin{table}[h]
\centering
\caption{ Numerical results of the Shannon entropy.}
\begin{tabular}{|c|c|c|c|c|c|}
\hline
$n$ & $a$ & $S_{x}$ & $S_{p}$ & $S_{x}+S_{p}$ & $1+ln\pi$ \\ \hline
 $0$ & $2$ & $\frac{1}{105}[638-105ln(280)]= 0.441401$ & $2.14029$ & $2.58169$ & $2.1447$\\
  & $4$ & $\frac{1}{105}[638-105ln(560)]= -0.251746$ & $2.83343$ & $2.58169$ & $2.1447$ \\
  & $6$ & $\frac{1}{105}[638-105ln(840)]= -0.657211$ & $3.2389$ & $2.58169$ & $2.1447$ \\ \hline
$1$ & $2$ & $0.582545$ & $2.7803$ & $3.36285$ & $2.1447$ \\
  & $4$ & $-0.110602$ & $3.47345$ & $3.36285$ & $2.1447$ \\
  & $6$ & $-0.516067$ & $3.87891$ & $3.36285$ & $2.1447$ \\ \hline
$2$ & $2$ & $0.647182$ & $3.14459$ & $3.79177$ & $2.1447$ \\
  & $4$ & $-0.0459656$ & $3.83773$ & $3.79177$ & $2.1447$ \\
  & $6$ & $-0.451431$ & $4.24322$ & $3.79177$ & $2.1447$\\ \hline
\end{tabular} 
\label{tab1}
\end{table}

\begin{figure}
\begin{center}
\begin{tabular}{ccc}
\includegraphics[height=5cm]{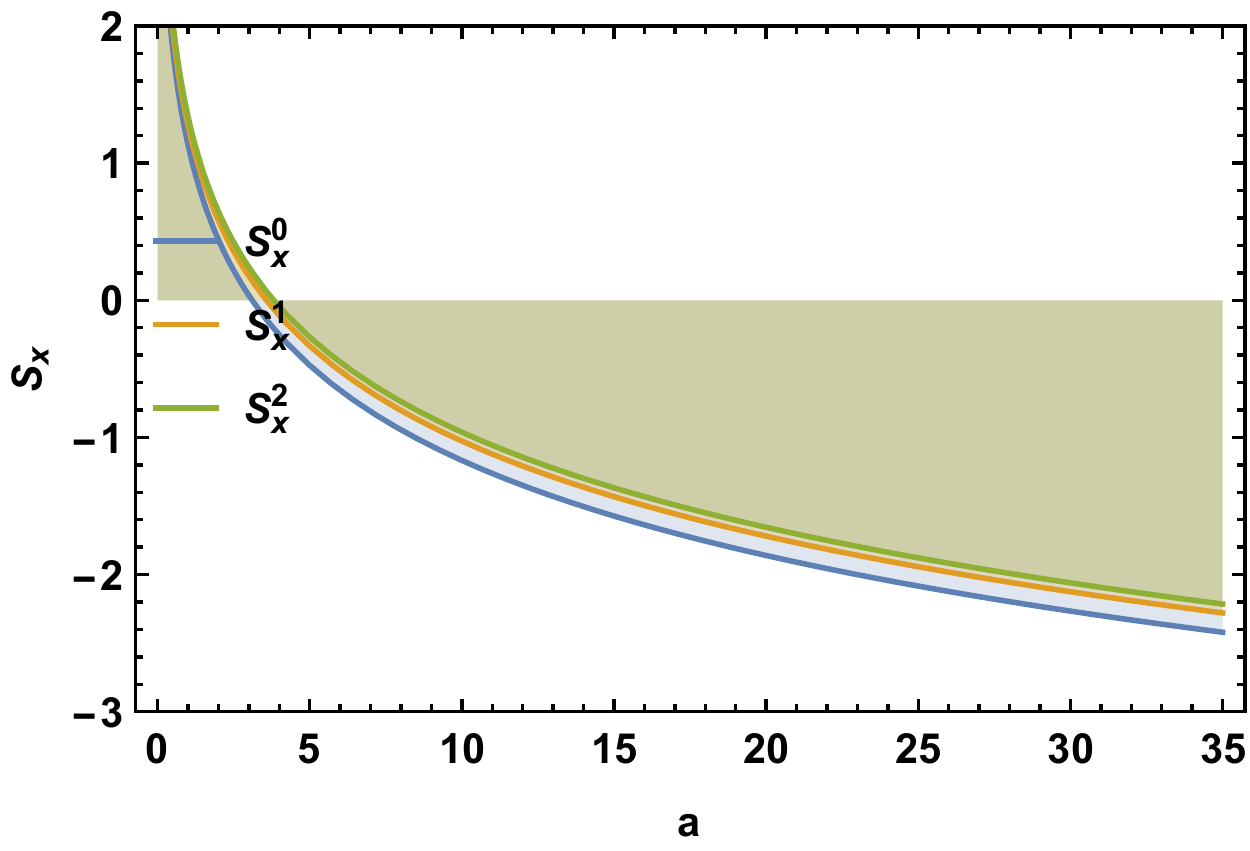} 
\includegraphics[height=5cm]{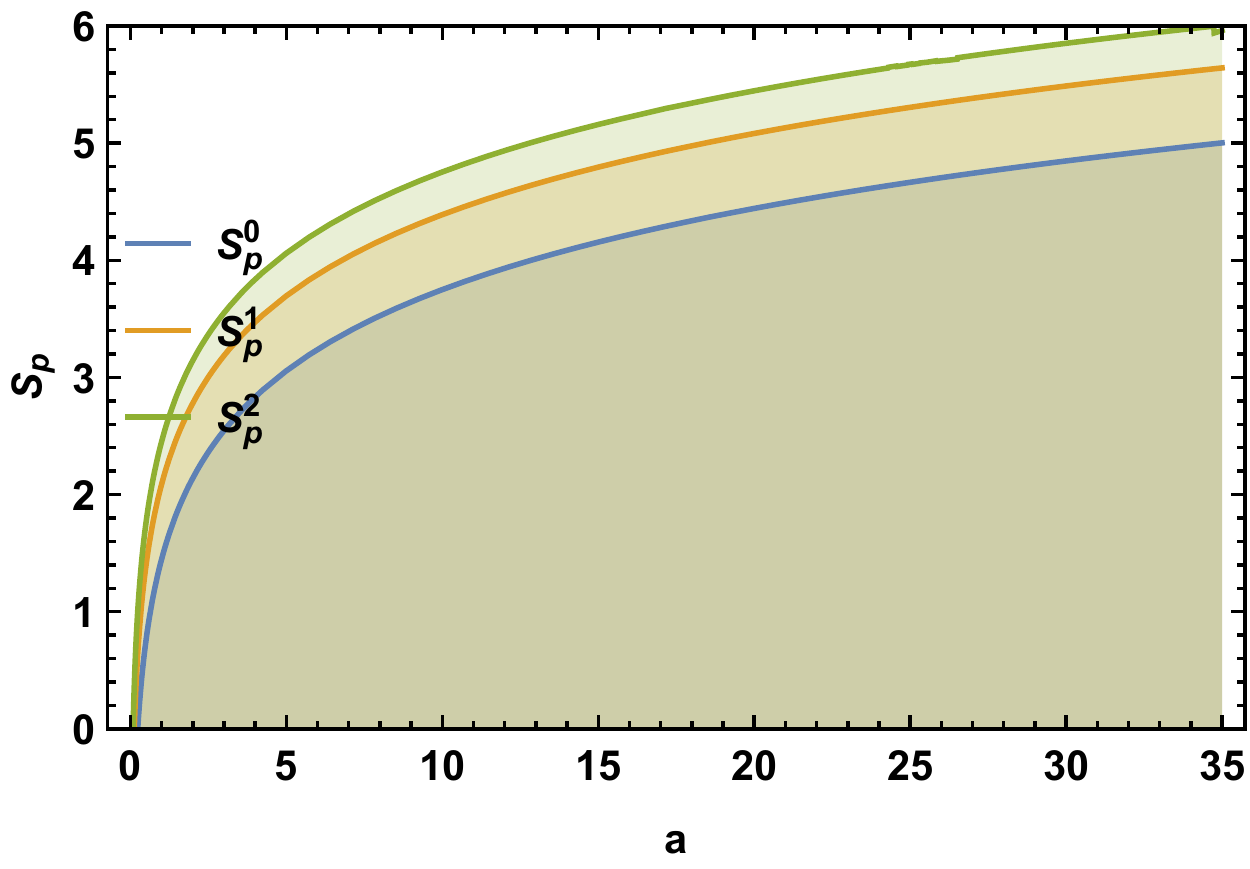}\\
(b) \hspace{6 cm}(c)
\end{tabular}
\end{center}
\caption{Plots of the Shannon entropy as function of the width of the mass distribution. (a) In position space. (b) In momentum space.
\label{fig9}}
\end{figure}

We can notice that the Shannon entropy tends to decrease in the space of the position $S_x$ according to the parameter $a$ that defines the width of the mass profile, while the entropy tends to increase proportionally in the momentum space $S_p$. With this, it can be seen that the quantity $S_{x}+S_{p}$ becomes invariant regardless of the parameter $a$ that defines the width of the spatial distribution of the solitonic mass. Figure \ref{fig9} depicts the behavior of $S_x$ and $S_p$ for the three lowest power levels in the system.

\subsection{Fisher's Theory}

In his 1925 work entitled \textit{``Theory of statistical estimation"} \cite{Fisher}, Ronald A. Fisher introduces his well-known theory as Fisher information which is a way of measuring the amount of information that an observable random variable carries concerning an unknown parameter \cite{Fisher}.

Fisher's information has drawn a lot of attention in several areas of computing, physics, and engineering, due to its numerous applications. For example, using Fisher's minimum information principle, one can obtain the non-relativistic quantum mechanics equations \cite{Frieden1,Reginatto}, the time independent Kohn-Sham equations and the time-dependent Euler equation \cite {Naje}. Another application that has gained supporters in recent years is the study of Fisher's information for position-dependent mass cases \cite{Falaye,Lima2021}, this is also due to the large applications of the concept of position-dependent mass in the quantum context \cite{Almeida,Almeida1,Almeida2}.

The interesting thing about this theory, very similar to Shannon's theory, is that it is related to the probability density of an information system, and it can be related to the probability density found in quantum systems $\vert\Psi(x )\vert^{2}$. An information density linked to Fisher's theory is defined to the position $\rho^n_F(x)$ and the moment $\rho^n_F(p)$ \cite{Falaye,Fisher}
\begin{eqnarray}
\rho^n_F(x)&=&\vert\Psi^n(x)\vert^{2}\bigg[\frac{d}{dx}\ln\vert\Psi^n(x)\vert^2\bigg]^2,\\
\rho^n_F(p)&=&\vert\Phi^n(p)\vert^{2}\bigg[\frac{d}{dp}\ln\vert\Phi^n(p)\vert^2\bigg]^2.
\end{eqnarray} 

Thus, for a probability density of a continuous system, Fisher's information can be defined for the position space $F^n_{x}$ and for the momentum space $F^n_{p}$ \cite{Fisher}
\begin{eqnarray}
F^n_{x}&=&\int_{-\infty}^{\infty}\vert\Psi^n(x)\vert^2\bigg[\frac{d}{dx}\ln\vert \Psi^n(x)\vert^2\bigg]^{2} dx > 0,\label{f11}\\
F^n_{p}&=&\int_{-\infty}^{\infty}\vert\Phi^n(p)\vert^2\bigg[\frac{d}{dp}\ln\vert\Phi^n(p)\vert^2\bigg]^{2} dp > 0\label{f12}.
\end{eqnarray}
For convenience, we can rewrite equation (\ref{f11}) in such a way that \cite{Falaye},
\begin{eqnarray}
F_{x}^{n}=4\int_{-\infty}^{\infty}\Psi_{n}(x)\Psi_{n}^{*'}(x)dx+\int_{-\infty}^{\infty}\bigg[\frac{\Psi_{n}^{'}(x)}{\Psi_{n}(x)}-\frac{\Psi_{n}^{*'}(x)}{\Psi_{n}^{*}(x)}\bigg]\vert\Psi_{n}(x)\vert^2 dx>0,
\end{eqnarray}
with the notation line ($'$) denoting the respective derivative of the wave function concerning variable $x$. Similarly, we can rewrite equation (\ref{f12}),
\begin{eqnarray}
F_{p}^{n}=4\int_{-\infty}^{\infty}\Phi_{n}(p)\dot{\Phi}_{n}^{*}(p)dp+\int_{-\infty}^{\infty}\bigg[\frac{\dot{\Phi}_{n}(p)}{\Phi_{n}(p)}-\frac{\dot{\Phi}_{n}^{*}(p)}{\Phi_{n}^{*}(p)}\bigg]\vert\Phi_{n}(p)\vert^2 dp>0.
\end{eqnarray}
where dot (\ $\dot{}$\ ) represents the derivative with respect to the independent variable $p$.

\subsubsection{Fisher's information Analysis}

\begin{figure}
\begin{center}
\begin{tabular}{ccc}
\includegraphics[height=5cm]{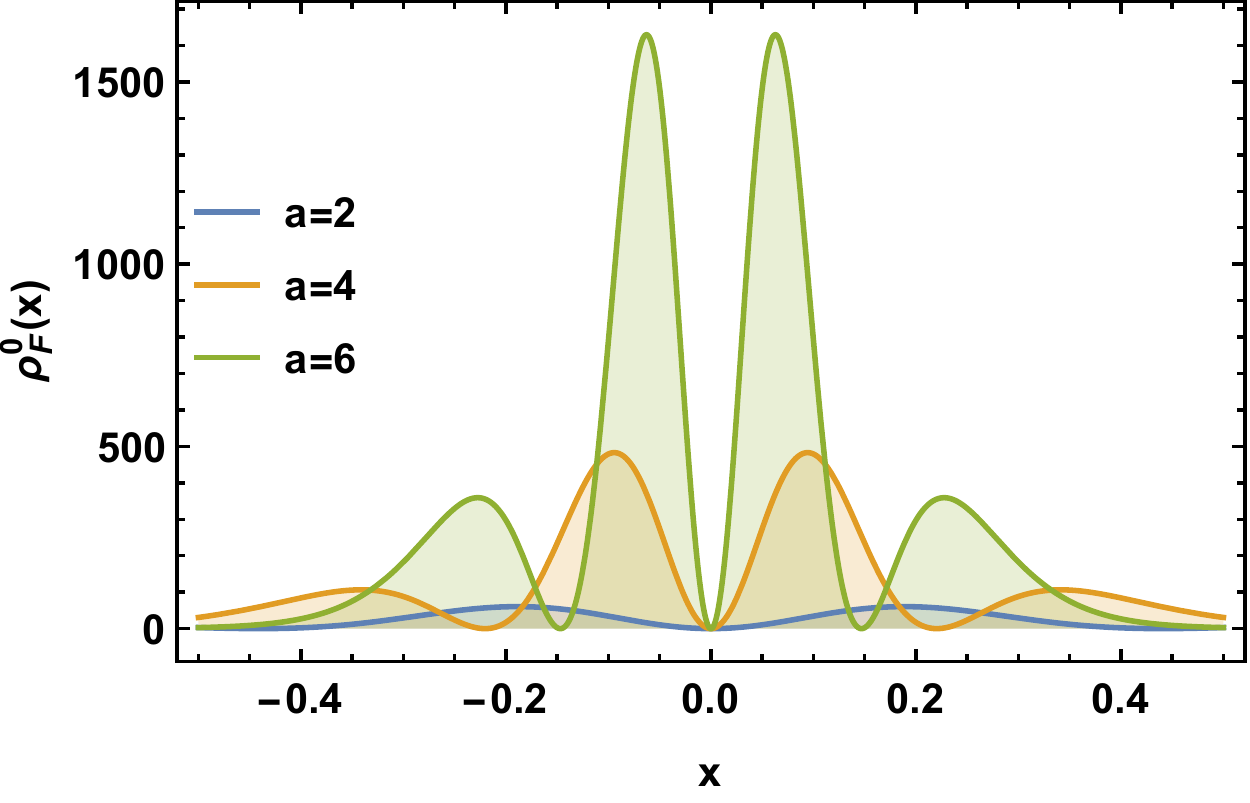}\\
(a)\\
\includegraphics[height=5cm]{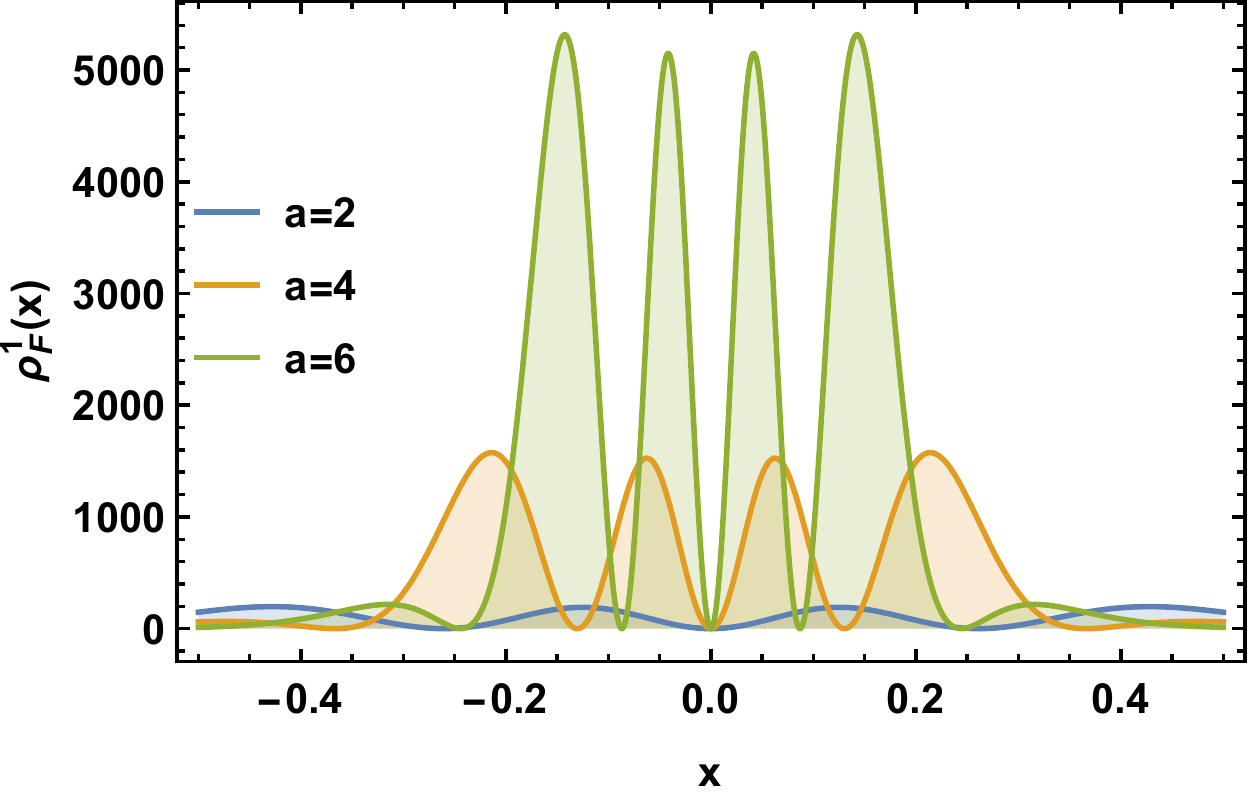} 
\includegraphics[height=5cm]{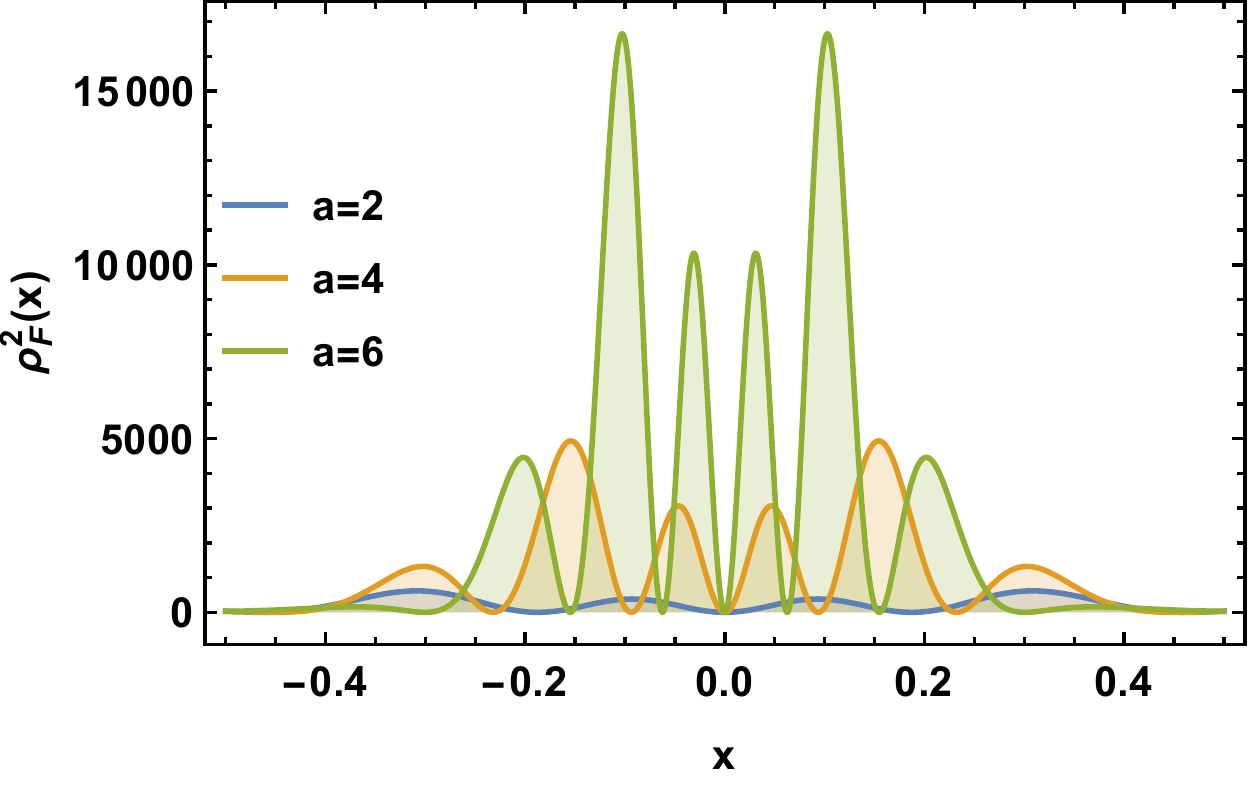}\\
(b) \hspace{6 cm}(c)
\end{tabular}
\end{center}
\caption{Behavior of information densities in position space $ \rho_F(p) $. (a) $n = 0$. (b) $n = 1$. (c) $n = 2$.
\label{fig10}}
\end{figure}

\begin{figure}
\begin{center}
\begin{tabular}{ccc}
\includegraphics[height=5cm]{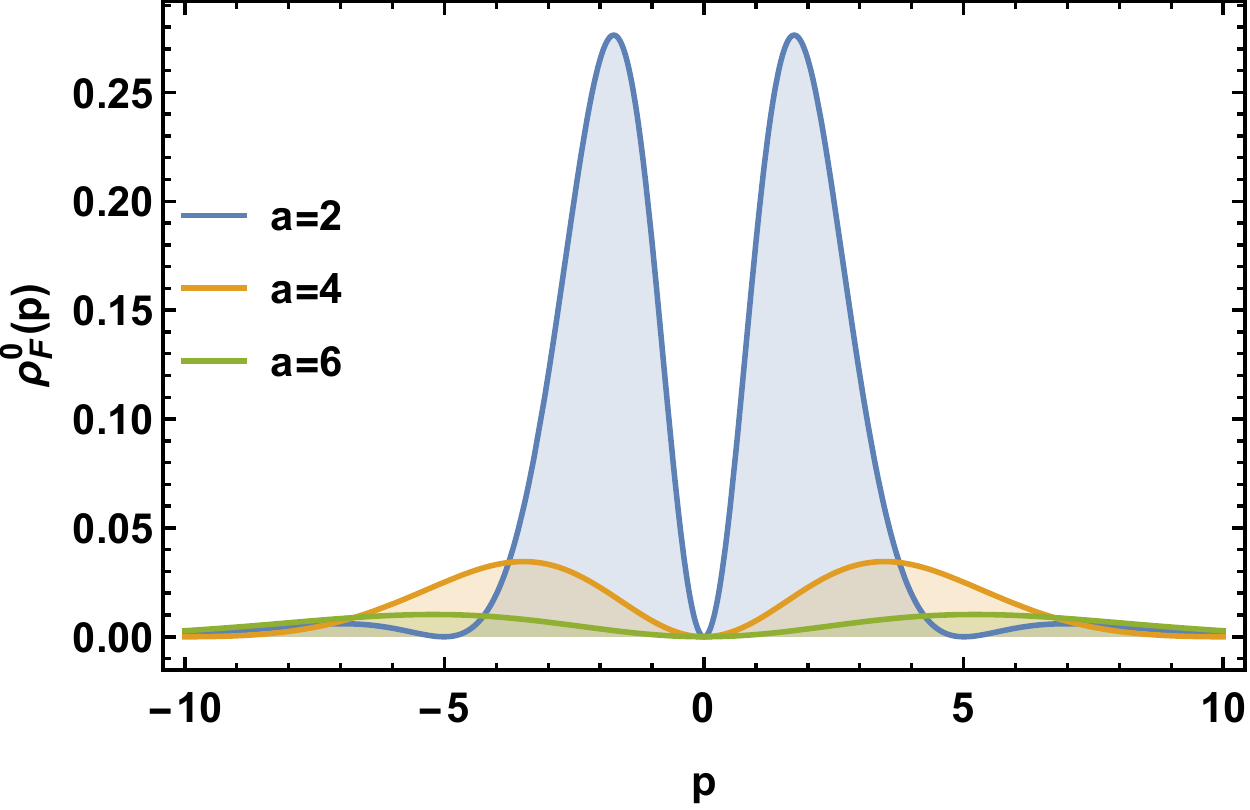}\\
(a)\\
\includegraphics[height=5cm]{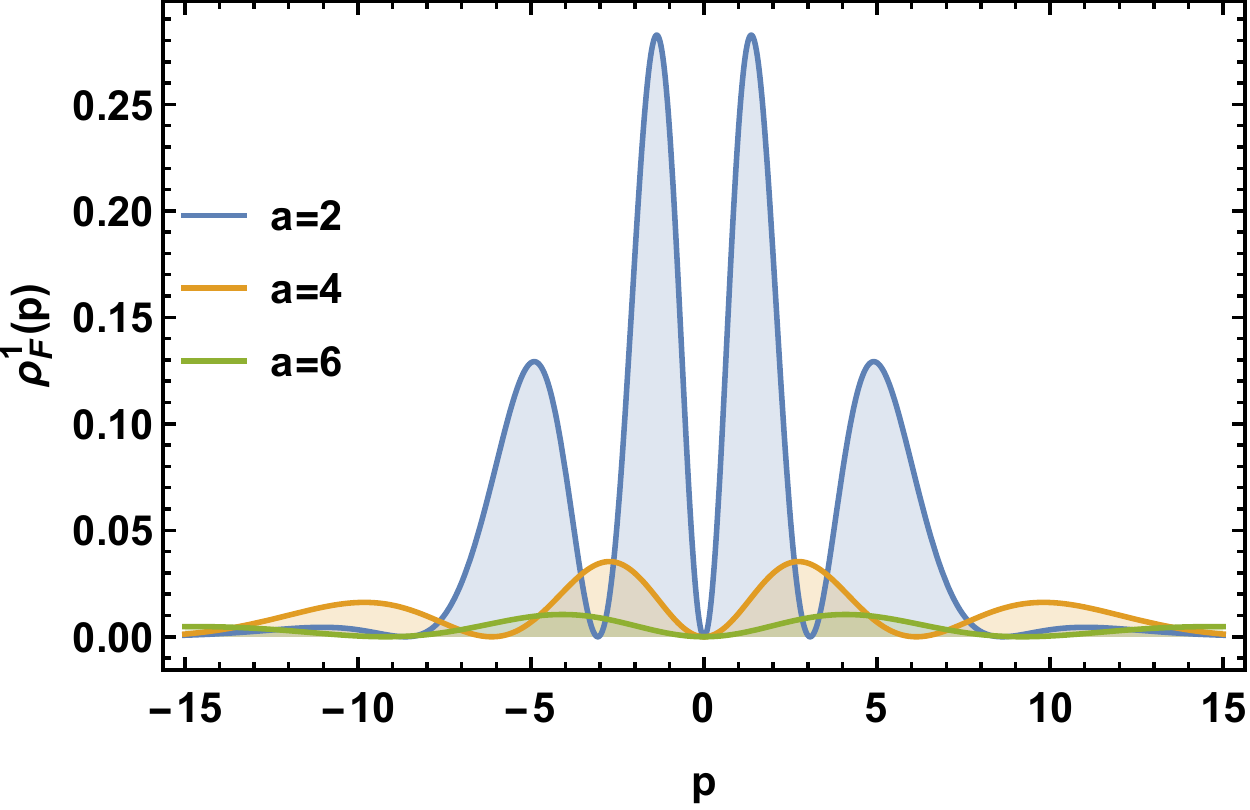} 
\includegraphics[height=5cm]{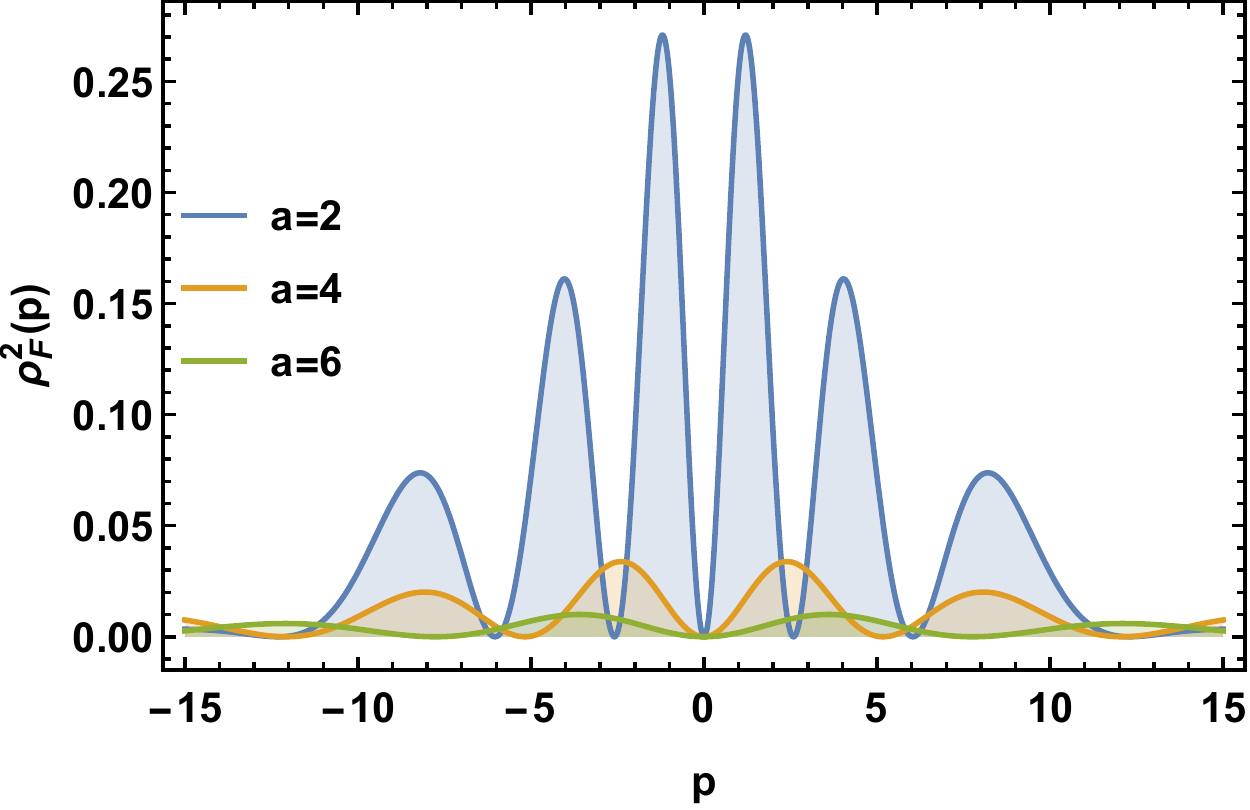}\\
(b) \hspace{6 cm}(c)
\end{tabular}
\end{center}
\caption{Behavior of information densities in momentum space $ \rho_F (p) $. (a) $n = 0$. (b) $n = 1$. (c) $n = 2$.
\label{fig11}}
\end{figure}

We apply Fisher's information to analyze solutions to our one-particle problem
with mass depending on the position in a hyperbolic barrier potential. To start with, it is interesting to observe the behavior of the information densities $\rho_F$, both with respect to the position space (\ref{solu}) and for the momentum space (\ref{solup}). In figure \ref{fig10} we plot the information densities in position space. The information densities in the momentum space are analyzed using the figure \ref{fig11}. With the table \ref{tab2}, the numerical study of Fisher's information was carried out considering the eigenfunctions in the space of position and moment, in addition to calculating the standard deviation of position and moment for these eigenfunctions
\begin{eqnarray}
\sigma_{x}^{2}&=&\langle x^{2}\rangle-\langle x\rangle^{2},\nonumber\\
\sigma_{p}^{2}&=&\langle p^{2}\rangle-\langle p\rangle^{2},
\end{eqnarray}
where $\langle x\rangle$, $\langle x^{2}\rangle$, $\langle p\rangle$ and $\langle p^{2}\rangle$ are the respective averages of the $x$ observables, $x^{2}$, $p$ and $p^{2}$  \cite{Griffiths,Cohen}. It is interesting to note that for all three energy states $n=0 ,1 , 2$, we obtain
\begin{equation}
\langle p\rangle=0 \qquad \mathrm{and} \qquad \langle x\rangle=0.
\end{equation}

\begin{table}[h]
\centering
\caption{{ Numerical results of the Fisher information.}}
\begin{tabular}{|c|c|c|c|c|c|c|c|c|}
\hline 
$n$ & $a$ & $<x^2>$ & $<p^2>$ & $\Delta x$ & $\Delta p$& $\Delta x \Delta p$ & $F_x$ & $F_p$\\
\hline
$0$ & $2$ & $0.302839 $&$ 8.88889$  & $0.550308$ & $2.98142$ & $1.6407$ & $35.5556$ & $1.21136$\\ 
  & $4$ &  $0.0757097 $& $35.55568$ & $0.275154$ & $5.96285$ &$ 1.6407 $& $142.222$ &$0.302839$ \\
  & $6$ & $0.0336488 $& $80.0$ & $0.183436$ & $8.94427$ & $1.6407 $& $320.0$ & $0.134595$\\ \hline
$1$ & $2$ &$0.3752$ & $34.188$ & $0.612536$ & $5.84705$ & $3.58153 $& $136.752$ &$ 1.5008$\\
  & $4$ & $0.0938 $& $136.752$ & $0.306268$ & $11.6941$ & $3.58153 $& $547.009$ &$0.3752$ \\
  & $6$ & $0.0416889 $& $307.692$ & $0.204179$ & $17.5412$ & $3.58153 $& $1230.77$ &$0.166756$ \\ \hline
$2$ & $2$ &$0.418019$ &$75.5113$ & $0.646544$ & $8.68972$ & $5.61829$ & $302.045$ &$1.67208$ \\
  & $4$ & $0.104505$ & $302.045$ & $0.323272$ & $17.3794$ &$ 5.61829 $& $1208.18$ &$0.418019$ \\
  & $6$ & $0.0464466$ & $679.602$ & $0.215515$ & $26.0692$ &$5.61829 $& $2718.41$ &$0.185786$ \\
\hline 
\end{tabular} 
\label{tab2}
\end{table}

\begin{figure}
\begin{center}
\begin{tabular}{ccc}
\includegraphics[height=5cm]{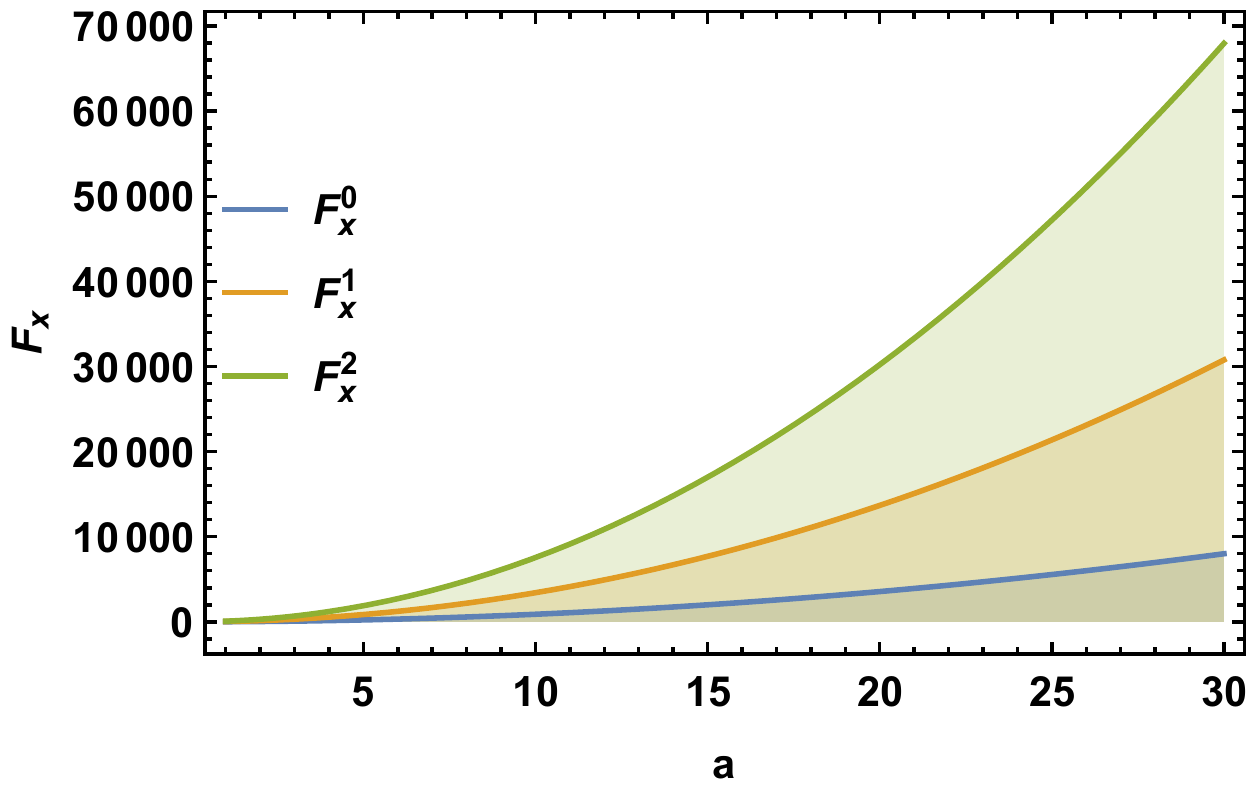} 
\includegraphics[height=5cm]{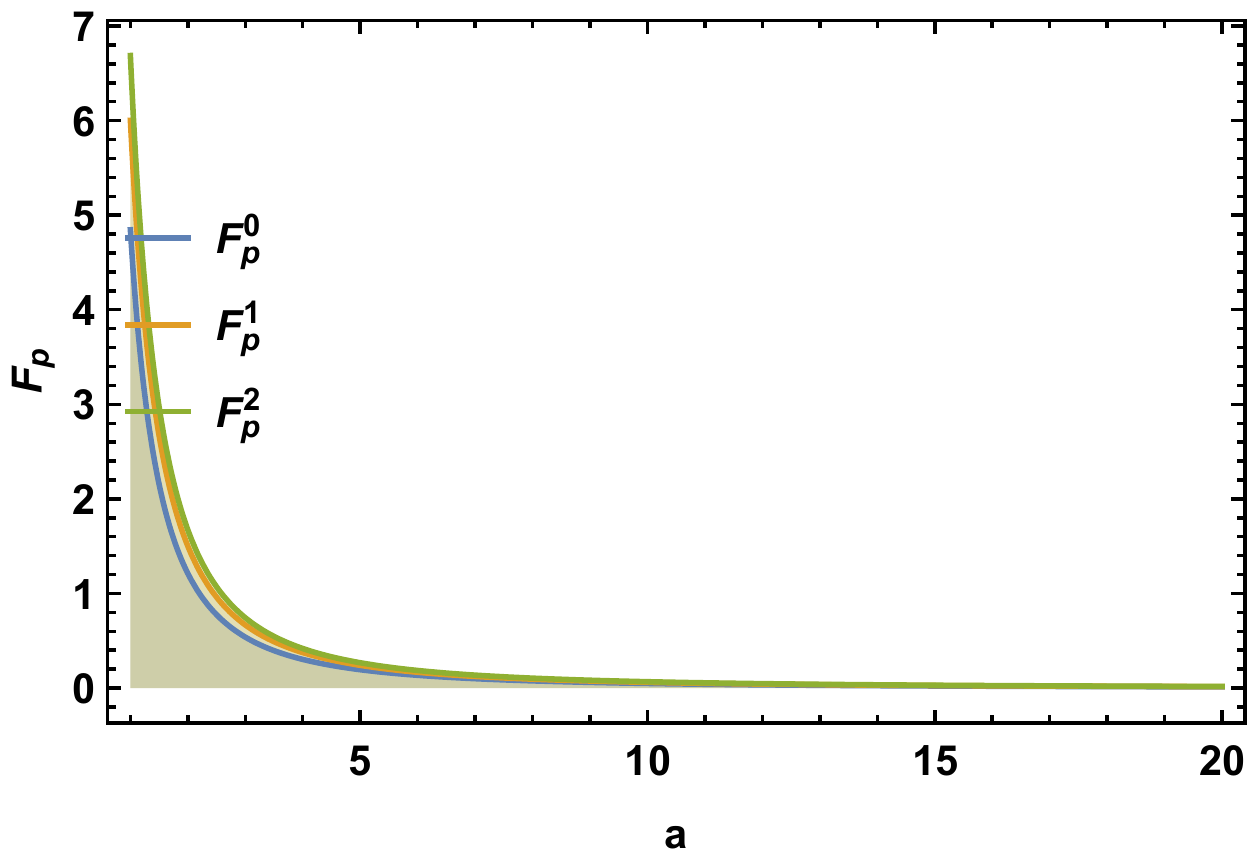}\\
(b) \hspace{6 cm}(c)
\end{tabular}
\end{center}
\caption{Plots of the Fisher information as function of the width of the mass distribution. (a) In position space. (b) In momentum space.
\label{fig12}}
\end{figure}

We note the existence of a ``propagation of information", where the Fisher information tends to increase with the width of the mass distribution, that is, in the space of positions the information increases proportionally with $a^{2}$. At first, the information tends to decrease with the width of the mass distribution, that is, it decreases with $a^{-2}$ in the space of momentum.We can observe this behavior more clearly in the figure \ref{fig12}.

We also observe with the values of the uncertainties $\Delta_x$ and $\Delta_p$ that the Heisenberg inequality is obeyed, obtaining the relations,
\begin{equation}
F_{x}\geq \frac{1}{(\Delta_x)^2}\qquad \mathrm{and} \qquad 
F_{p}\geq \frac{1}{(\Delta_p)^2}.
\end{equation}

In this way, we can write the Heisenberg uncertainty principle as 
\begin{equation}
\sigma_{x}\sigma_{p}\geq\frac{1}{(F_{x}F_{p})^{\frac{1}{2}}}\geq\frac{\hbar}{2},
\end{equation}
that is
\begin{equation}
F_{x}F_{p}\geq4\hbar^{-2}.
\end{equation}

\section{Final remarks}
\label{sec3}

In this work, we present the solutions of the Schrödinger equation for a solitonic mass distribution with the kinetic energy operator ordered by BenDaniel-Duke when subjected to barrier potentials $V(x)=V_1\coth^2(x)+V_2 \mathrm{csch}^2(x)$. We find the eigenfunctions and the corresponding quantized energies. With this, it was possible to observe that the complete set of solutions that describe these physical systems are given by the known hypergeometric functions of Gauss. We also present, through the Fourier transform, the eigenfunctions for the momentum space.

With the analytical solutions of the two systems studied, we calculate the Shannon entropy for the first energy levels, both for the position space $S_x$ and for the momentum space $S_p$. With this, we can conclude that the Shannon entropy tends to decrease with the width of the mass distribution in the position space, on the other hand, the Shannon entropy tends to increase for the momentum space, so that it makes the sum $S_{x}+S_{p}$ be constant regardless of the width of the soliton. With this, we note that the BBM relationship was followed for both analyzed cases.

Finally, with the help of the solutions found, we calculate Fisher information for the first energy levels of each case, both for the space of positions $F_x$ and for the space of moments $F_p$. With this, we observe a behavior contrary to that found in the Shannon entropy, where the Fisher information tends to increase in the position space. In contrast, Fisher's information decreases in momentum space. We also observe that $\Delta_x\Delta_p$ is constant with the width of the mass profile for both cases. With this, we conclude that the uncertainty of the measurements of the observables tends to be minimum in the position space and maximum in the momentum space, thus respecting the Heisenberg uncertainty principle. We also obtain that Fisher's information is related to the uncertainties of position and momentum, that is, $F_{x}F_{p}\geq4\hbar^{-2}$. Therefore, we can conclude that the more localized the mass distribution, the more information transmitted.

\section*{Acknowledgments}
The author thank the Coordena\c{c}\~{a}o de Aperfei\c{c}oamento de Pessoal de N\'{i}vel Superior (CAPES) for financial support. The author thanks C. A. S. Almeida and F. C. E. Lima for the important discussion, and is grateful to M. S. Cunha for the valuable discussion and important contribution to the progress of this project.


\begin{thebibliography}{99}

\bibitem{Schr}
E. Schr\"{o}dinger, Phys. Rev. \textbf{28}, 1049 (1926).

\bibitem{Griffiths}
D. J. Griffiths, {\it Introduction to quantum mechanics}. Pearson International Edition (Pearson Prentice Hall, Upper Saddle River, 2005), (1960).

\bibitem{Roos}
O. von Roos, Phys. Rev. B \textbf{27}, 7547 (1983).

\bibitem{Bastard}
G. Bastard, {\it Wave mechanics applied to semiconductor heterostructures}. Les editions
de Physique, Les Ulis, Fran\c{c}a, v. 66, 042116, p.707, 1988.

\bibitem{Weisbuch}
C. Weisbuch, and B. Vinter, {\it Quantum semiconductor heterostructures}. New York:
Academic Press, (1993).

\bibitem{Luttinger}
J. M. Luttinger, and W. Kohn, Phys. Rev. {\bf 97}, 869 (1955).

\bibitem{Nabu}
R. A. El-Nabulsi, Physica E: Low-dimensional Systems and Nanostructures {\bf 127}, 114525, (2021).

\bibitem{Pourali}
B. Pourali, B. Lari, and H. Hassanabadi, Physica A: Statistical Mechanics and its Applications {\bf 584}, 126374 (2021).

\bibitem{Kasapoglu}
E. B. Al, E. Kasapoglu, S. Sakiroglu, H. Sari, and I. S\"{o}kmen, Materials Science in Semiconductor Processing \textbf{135}, 106076 (2021).

\bibitem{Mustafa}
O. Mustafa, and S. H. Mazharimousavi, Int. J. Ther. Phys.  \textbf{47}, 446 (2008).

\bibitem{Sever}
R. Sever, and C. Tezcan, Int. J. Mod. Phys. E {\bf 17}, 1327 (2008).

\bibitem{Plastino}
A. R. Plastino, A. Rigo, M. Casas, F. Garcias, and  A. Plastino, Phys. Rev. A {\bf 60}, 4318 (1999).

\bibitem{Almeida}
F. S. A. Cavalcante, R. N. C. Filho, J. R. Filho, C. A. S. Almeida, and V. N. Freire,  Phys. Rev. B {\bf 55}, 1326 (1997).

\bibitem{Almeida1}
R. Renan, M. H. Pacheco and C. A. S. Almeida, J. Phys. A: Math. Gen. {\bf 33}, L509 (2000).

\bibitem{Almeida2}
A. S. Dutra, and C. A. S. Almeida, Phys. Lett. A {\bf 275}, 25 (2000).

\bibitem{Cunha}
M. S. Cunha, and H. R. Christiansen, Comm. Theor. Phys. \textbf{60}, 642 (2013).

\bibitem{Dong}
J. Yu, S. H. Dong, and  G. H. Sun, Phys. Lett. A \textbf{322}, 290 (2004).

\bibitem{Dong1}
S. H. Dong, and M. Lozada-Cassou, Phys. Lett. A, \textbf{337}, 313 (2005).

\bibitem{Falaye}
B.J. Falaye,  F. A. Serrano, and S. H. Dong, Phys. Lett. A \textbf{380}, 267 (2016).

\bibitem{Macedo2015}
D.X. Macedo, I. Guedes, Phys. A \textbf{434}, 211–219 (2015). 

\bibitem{Lima2021}
F. C. E. Lima,
 `` Quantum information entropies for a soliton at hyperbolic well ''
\url{https://arxiv.org/abs/2110.11195}.  

\bibitem{Sun2015}
G. H. Sun, P. Dušan,  C. N. Oscar, S. H. Dong,  Chin. Phys. B \textbf{24}, 100303 (2015).

\bibitem{Dong2016}
S. Dong, G. H. Sun, B. J. Falaye, S. H. Dong,  Eur. Phys. J. Plus  \textbf{131}, 176 (2016).

\bibitem{Navarro}
G. Ya\~{n}ez-Navarro, G. H. Sun, T. Dytrych, K. D. Launey, S. H. Dong, and J. P. Draayer, Ann. Phys. {\bf 348}, 153 (2014).

\bibitem{Shannon}
C. E. Shannon, The Bell System Tecnical Journal {\bf 27}, 379 (1948).

\bibitem{Kripp}
K. Krippendorff, Mathematical theory of communication, Departmental Pepers (ASC), 169 (2009).

\bibitem{Zhou}
M. Zhou, {\it AI Uncertainty Based on Rademacher Complexity and Shannon Entropy}, arXiv:2102.07638, (2021).

\bibitem{Grasselli}
F. Grasselli, {\it Elements of Quantum Information Theory}. In: Quantum Cryptography, Springer, Cham., (2021).

\bibitem{Amigo}
J. M. Amig\'{o}, R. Dale, and P. Tempesta, Chaos \textbf{31}, 013115 (2021).

\bibitem{Lima}
F. C. E. Lima, A. R. P. Moreira, and C. A. S. Almeida, Int. J. Quant. Chem. \textbf{121}, e26645 (2021).

\bibitem{Lima1}
F. C. E. Lima, A. R. P. Moreira, L. E. S. Machado, and C. A. S. Almeida, Int. J. Quant. Chem. \textbf{121}, e26749 (2021).

\bibitem{Fisher}
 R. A. Fisher, Math. Proc. Camb. Philos. Soc. {\bf 22}, 700 (1925).

\bibitem{Shi}
Y. J. Shi, G. H. Sun, J. Jing, and S. H. Dong, Laser Phys. \textbf{27}, 125201 (2017).

\bibitem{Arenas}
A. J. Torres-Arenas, Q. Dong, G. H. Sun, and S. H. Dong, Phys. Lett. A \textbf{382}, 1752 (2018).

\bibitem{Ikot}
A. N. Ikot, G. J. Rampho, P. O. Amadi, M. J. Sithole, U. S. Okorie, and M. I. Lekala, Europ. Phys. J. Plus {\bf 135}, 1-13 (2020).

\bibitem{BenDaniel}
D. J. BenDaniel, and C. B. Duke, Phys. Rev. A {\bf 152}, 683 (1966).

\bibitem{GoraW}
T. Gora, and F. Williams, Phys. Rev. {\bf 177}, 1179 (1969).

\bibitem{ZhuKroemer}
Q. G. Zhu, and H. Kroemer, Phys. Rev. B {\bf 27}, 3519 (1983).

\bibitem{LiKuhn}
L. T. Li, and K. J. Kuhn, Phys. Rev. B {\bf 47}, 12760 (1993).

\bibitem{Bagchi}
B. Bagchi, P. Gorain, C. Quesne, R. Roychoudhury, Phys.
Lett. A {\bf 19}, 2765 (2004).

\bibitem{Heeger}
A. J. Heeger, K. Sivelson, J. R. Schrieffer, and W. P. Su,  Rev. Mod. Phys. {\bf 60}, 781 (1988).

\bibitem{Kartashov}
Y. V. Kartashov, V. A. Vysloukh, and L.Torner, Phys. Rev. Lett. {\bf 96}, 073901 (2006).

\bibitem{Rajaraman}
R. Rajaraman, {\it Solitons and instantons}, (1982).

\bibitem{Nalewajski}
R. F. Nalewajski, Int. J. Quant. Chem. {\bf 108}, 2230 (2008).

\bibitem{Nagaoka}
H. Nagaoka, In Asymptotic Theory Of Quantum Statistical Inference: Selected Papers, 113 (2005).

\bibitem{Wang}
B. Wang, D. Zhao, T. Lu, S. Liu, and C. Rong, J. Phys. Chem. A {\bf 125}, 3802 (2021).

\bibitem{Lian}
Yi-Jun Lian, and Jin-Ming Liu. Comm. Theor. Phys. {\bf 73}, 085102 (2021).

\bibitem{Falaye0}
B. J. Falaye, and M. S. Liman, Laser Phys. \textbf{30}, 225206 (2020).

\bibitem{Rothstein}
J. Rothstein,  Science {\bf 114} 171 (1951).

\bibitem{Zou}
N. Zou, J. Phys.: Conf. Ser. {\bf 1827}, 012120 (2021).

\bibitem{Serrano}
F. A. Serrano, B. J. Falaye, S. H. Dong, Phys. A \textbf{446}, 152 (2016).

\bibitem{Burke}
K. Burke, J. Werschnik, E. K. U. Gross, J. Chem. Phys. {\bf 123}, 62206 (2005).

\bibitem{Gadre1985}
S.~R.~Gadre, S.~B.~Sears, S.~J.~Chakravorty and R.~D.~Bendale,
Phys. Rev. A \textbf{32}, 2602-2606 (1985).

\bibitem{Frieden1}
B. R. Frieden, J. Mod. Opt. {\bf 35}, 1297 (1988).

\bibitem{Callen}
Herbert B. Callen, {\it Thermodynamics and an Introduction to Thermostatistics}, 2nd ed., Wiley, New York, (1985).

\bibitem{Pathria}
R. K. Pathria, and P. D. Beale, {\it Statistical mechanics}, 2nd ed., Butterworth-Heinemann, (1996).

\bibitem{Born}
M. Born, Science {\bf 122}, 675 (1955).

\bibitem{Hirschmann}
I. I. Hirschmann Jr., Amer. J. Math.  {\bf 79}, 152 (1957).

\bibitem{Beckner}
W. Beckner, Ann. of Math. {\bf 102}, 159 (1975).

\bibitem{Bialy}
I. Bialynicki-Birula, and  J. Mycielski, Commun. Math. Phys. {\bf 44}, 129 (1975).

\bibitem{Reginatto}
M. Reginatto, Phys. Rev. A {\bf 58}, 1775 (1998).

\bibitem{Naje}
R. F. Nalewajski, Chem. Phys. Lett. {\bf 372}, 28 (2003).

\bibitem{Cohen}
Claude Cohen-Tannoudji, B. Diu, F. Lalo\"{e}, {\it Quantum Mechanics Volume I: Basic Concepts, Tools, and Applications}. 2nd. ed., John Wiley $\&$ Sons, (2019).


\end{thebibliography}
\end{document}